\documentclass[english,reprint, citeautoscript, aip, jcp]{revtex4-1}
\usepackage[T1]{fontenc}
\usepackage[utf8]{inputenc}
\usepackage{verbatim}
\usepackage{amsmath}
\usepackage{amssymb}
\usepackage{graphicx}

\makeatletter
\usepackage{algorithm,algpseudocode}

\usepackage{color}
\newcommand{\kb}[1]{\textcolor{red}{[XX #1 XX]}}

\makeatother

\usepackage{babel}
\begin{document}

\title{Gradient-based stochastic estimation of the density matrix}

\author{Zhentao Wang}

\affiliation{Department of Physics and Astronomy, The University of Tennessee,
Knoxville, TN 37996, USA}

\author{Gia-Wei Chern}

\affiliation{Department of Physics, University of Virginia, Charlottesville, VA
22904, USA}

\author{Cristian D. Batista}

\affiliation{Department of Physics and Astronomy, The University of Tennessee,
Knoxville, TN 37996, USA}

\affiliation{Quantum Condensed Matter Division and Shull-Wollan Center, Oak Ridge
National Laboratory, Oak Ridge, TN 37831, USA}

\author{Kipton Barros}
\email{kbarros@lanl.gov}

\affiliation{Theoretical Division and CNLS, Los Alamos National Laboratory, Los
Alamos, NM 87545, USA}
\begin{abstract}
Fast estimation of the single-particle density matrix is key to many
applications in quantum chemistry and condensed matter physics. The
best numerical methods leverage the fact that the density matrix elements
$f(H)_{ij}$ decay rapidly with distance $r_{ij}$ between orbitals.
This decay is usually exponential. However, for the special case of
metals at zero temperature, algebraic decay of the density matrix
appears and poses a significant numerical challenge. We introduce
a gradient-based probing method to estimate all local density matrix
elements at a computational cost that scales linearly with system
size. For zero-temperature metals the stochastic error scales like
$S^{-(d+2)/2d}$, where $d$ is the dimension and $S$ is a prefactor
to the computational cost. The convergence becomes exponential if
the system is at finite temperature or is insulating.
\end{abstract}
\maketitle
\global\long\def\d{\mathrm{d}}
\global\long\def\tr{\mathrm{\mathrm{tr}}\,}

\section{Introduction}

Many topics in quantum chemistry and condensed matter physics involve
an effective Hamiltonian,
\begin{equation}
\mathcal{\hat{H}}=\sum_{i,j=1}^{N}c_{i}^{\dagger}H_{ij}c_{j},\label{eq:hamiltonian}
\end{equation}
which is quadratic in fermionic creation and annihilation operators
($c_{i}^{\dagger}$ and $c_{i}$). Examples include density-functional
tight-binding models~\cite{Elstner98,Elstner14} for molecular dynamics
simulation~\cite{Aradi15}, Kondo lattice models of itinerant electrons
interacting with localized magnetic moments~\cite{Doniach77}, Falicov-Kimball
models of metal-insulator transitions~\cite{Falicov69}, and Bogoliubov-de
Gennes equations for superconductivity~\cite{Gennes66}. The index
$i$ specifies a single-particle wave function (position, spin, orbital
index, etc.). The electronic free energy in the grand canonical ensemble
is given by $\Omega=-k_{B}T\ln Z$, where $k_{B}$ is the Boltzmann
constant and $T$ is the temperature. The partition function is $Z=\tr e^{-\beta(\mathcal{\hat{H}}-\mu\hat{N}_{e})}$,
with $\beta=1/k_{B}T$, chemical potential $\mu$, and electron number
$\hat{N}_{e}=\sum_{i=1}^{N}c_{i}^{\dagger}c_{i}$. Evaluating the
above trace over fermions yields
\begin{equation}
\Omega=\sum_{\epsilon}g(\epsilon)=\tr g(H),\label{eq:free_energy}
\end{equation}
where $\{\epsilon\}$ are the eigenvalues of the single-particle Hamiltonian
matrix $H$, and
\begin{equation}
g(x)=-\beta^{-1}\ln[1+e^{-\beta(x-\mu)}].\label{eq:en_func}
\end{equation}
Note that the derivative of $g(x)$ is the usual Fermi function,
\begin{equation}
\frac{\d g(x)}{\d x}=f(x)=\frac{1}{e^{\beta(x-\mu)}+1}.\label{eq:fermi}
\end{equation}
Consequently, the derivative of the free energy gives the density
matrix,
\begin{equation}
f(H)=\d\Omega/\d H^{T}.\label{eq:density_matrix}
\end{equation}
Density matrix elements represent two-body correlations, $f(H)_{ij}=\langle c_{j}^{\dagger}c_{i}\rangle$.
Diagonal elements $f(H)_{ii}$ give the charge localized at $i$.
The expected electron number $\langle\hat{N}_{e}\rangle$ is 
\begin{equation}
N_{e}=\tr f(H).\label{eq:N_e}
\end{equation}

Efficient estimation of the density matrix, especially for metals,
is the central topic of this paper. Our primary motivation is to enable
dynamical simulations of effectively classical degrees of freedom
$\{\mathbf{x}_{1,}\mathbf{x}_{2}\dots\}$. For example, in quantum
molecular dynamics, $\mathbf{x}_{\alpha}$ may be positions of nuclei
evolving classically under the Born Oppenheimer approximation~\cite{Goedecker94,Voter96,VandeVondele12,Cawkwell12,Chern17a}.
In applications to itinerant magnets, $\mathbf{x}_{\alpha}$ may represent
a field of local moments~\cite{Barros13,Barros14b,Ozawa16,Wang16,Ozawa17,Ozawa17a,Chern18}.
The $N\times N$ single-particle Hamiltonian $H$ evolves with the
dynamical variables $\mathbf{x}_{\alpha}$. The electronic free energy
$\Omega$ may be augmented with classical interactions solely involving
the $\mathbf{x}_{\alpha}$; such interactions are straightforward
to handle, and we ignore them here.

The dynamics of $\mathbf{x}_{\alpha}$ is driven by effective forces
associated with energy derivatives. Referring to Eq.~(\ref{eq:density_matrix}),
the chain rule yields 
\begin{equation}
-\frac{\partial\Omega}{\partial\mathbf{x}_{\alpha}}=-\sum_{ij}f(H)_{ji}\frac{\partial H_{ij}}{\partial\mathbf{x}_{\alpha}}.\label{eq:density_chain}
\end{equation}
The matrix $\partial H/\partial\mathbf{x}_{\alpha}$, for each $\alpha$,
is typically highly localized and easy to compute. At every dynamical
time-step, a key numerical challenge is to calculate density matrix
elements $f(H)_{ij}$ for nearby states $i$ and $j$.

Direct diagonalization of the single-particle Hamiltonian $H$ is
possible but the $\mathcal{O}(N^{3})$ cost would severely limit system
sizes. Better methods take advantage of the sparsity of $H$. In a
real-space basis, the elements $H_{ij}$ typically decay exponentially
with spatial distance $r_{ij}=|\mathbf{r}_{i}-\mathbf{r}_{j}|$. %
{} If the system is either insulating or at finite temperature, then
the density matrix $f(H)_{ij}$ \emph{also} decays exponentially in
$r_{ij}$~\cite{Kohn59}. A rich set of algorithms have emerged to
calculate $f(H)$ by leveraging its sparsity~\cite{Goedecker99,Bowler12}.
Methods based upon iterated self-multiplication of sparse matrices~\cite{Beylkin99,Niklasson02,Niklasson03}
enable quantum molecular dynamics simulations with up to millions
of atoms~\cite{Cawkwell12,VandeVondele12}.

In the case of metals at zero temperature, however, $f(H)_{ij}$ decays
just algebraically in $r_{ij}$. Consequently, state-of-the-art methods
based upon sparse matrix-matrix multiplication are infeasible. Here
we consider instead stochastic methods that require only sparse matrix-vector
multiplication. 

In a direct probing approach, one may approximate $f(H)\approx[f(H)R]R^{\dagger}$~\cite{Curtis74,Coleman83}.
The random matrix $R$ contains $N\times S$ elements. The parameter
$S$ becomes a prefactor to the computational cost and controls accuracy.
Although never explicitly constructed, the outer product matrix $RR^{\dagger}$
is an unbiased approximation to the $N\times N$ identity matrix.
References~\onlinecite{Bekas07,Tang12} introduce a coloring strategy
to design $R$ to best leverage the spatial decay of $f(H)$. With
this strategy, we show that the stochastic error for direct probing
scales like $\Delta f\sim S^{-(d+1)/2d}$ for bulk $d$-dimensional
metals at zero temperature.

Inspired by Eq.~(\ref{eq:density_matrix}) and the favorable decay
properties of $g(H)$, we introduce a gradient-based probing approximation,
$f(H)\approx(\d/\d H^{T})\tr g(H)RR^{\dagger}$, and show that its
error scales like $\Delta f\sim S^{-(d+2)/2d}$. This approximation
scheme and its rapid convergence are our main results.

Crucially, the accuracy in estimating density matrix elements is independent
of system size $N$. Our probing method thus enables truly linear-scaling
dynamical simulations of metals. Furthermore, to a first approximation,
the unbiased stochastic errors in the forces {[}cf. Eq.~(\ref{eq:density_chain}){]}
can be absorbed into the noise term of a Langevin thermostat~\cite{Krajewski06,Barros13,Luo14,Arnon17}.
Alternatively, for fixed matrix $R$, gradient-based probing yields
conservative forces that generate time-reversible dynamics. Empirically,
we commonly find that $S\lesssim100$ random column vectors enable
accurate dynamical simulations over a wide range of temperatures.

\section{Stochastic trace estimation\label{sec:traces}}

We begin with stochastic estimation of the matrix traces, Eqs.~(\ref{eq:N_e})
and~(\ref{eq:free_energy}). One may approximate the electron number
as
\begin{equation}
N_{e}=\tr f(H)\approx\sum_{s=1}^{S}r^{(s)\dagger}f(H)r^{(s)}=\tr R^{\dagger}f(H)R,\label{eq:probing_Ne}
\end{equation}
where $R$ is a stochastic, $N\times S$ matrix composed of column
vectors $r^{(s)}$. Typically, $S\ll N$. The approximation error
is~\cite{Bekas07}

\begin{equation}
\Delta N_{e}=\tr f(H)(RR^{\dagger}-I).\label{eq:probing_err}
\end{equation}
Observe that the approximation is unbiased, $\langle\Delta N_{e}\rangle=0$,
provided that, on average, $\langle RR^{\dagger}\rangle=I$. The quality
of the approximation will typically improve with the number of column
vectors $S$.

The free energy may be approximated similarly,
\begin{equation}
\Omega\approx\tr R^{\dagger}g(H)R.\label{eq:probing_omega}
\end{equation}
The free energy error analysis is completely analogous to that of
$\Delta N_{e}$, which we will present below.

A remark on our numerical implementation: The matrix-vector products
$f(H)r^{(s)}$ and $g(H)r^{(s)}$ can be approximated at a cost that
scales linearly with system size $N$. In our approach we expand $f(H)$
and $g(H)$ in Chebyshev polynomials over $H$ using the Kernel Polynomial
Method (Appendix~\ref{sec:kpm})~\cite{Silver94,Silver96,Weisse06};
this method is simple and amenable to a gradient transformation (Appendix~\ref{sec:autodiff})~\cite{Griewank89,Barros13}.
Other methods, such as rational approximation~\cite{Ceriotti08,Sidje11},
are also possible, at least in principle. Alternative trace estimators
have also been proposed~\cite{Lin16,Lin17}. 

By numerically inverting the approximation of Eq.~(\ref{eq:probing_Ne}),
we can allow the chemical potential $\mu$ to vary according to a
fixed electron number $N_{e}$. Within this canonical ensemble, the
density matrix $f(H)$ may still be interpreted as the gradient of
the characteristic free energy. Appendix~\ref{sec:canonical} presents
the details of this transformation.

\subsection{Uncorrelated probing}

One possible choice for the stochastic matrix $R$ is

\begin{equation}
R_{is}=\zeta_{is}/\sqrt{S},\label{eq:R_uncorrelated}
\end{equation}
where $\zeta_{is}$ are uncorrelated random numbers that satisfy $\langle\zeta_{is}\zeta_{jt}^{\ast}\rangle=\delta_{ij}\delta_{st}$.
It is advantageous to constrain $|\zeta_{is}|=1$ such that $(RR^{\dagger})_{ii}=1$
and only off-diagonal elements contribute to the error $\Delta N_{e}$.
If we select $\zeta_{is}$ to be complex numbers with uniformly random
phases~\cite{Iitaka04} then one can show that the variance of the
stochastic error is
\begin{equation}
\mathrm{var}[\Delta N_{e}]=\frac{1}{S}\sum_{i<j}|f(H)_{ij}|^{2}\quad\;(R\,\mathrm{uncorrelated)}.\label{eq:var_uncorrelated}
\end{equation}
Observe that approximation~(\ref{eq:probing_Ne}) implicitly benefits
from the smallness of the off-diagonal elements of $f(H)$.

The idea to estimate traces using uncorrelated random column vectors
appeared in Refs.~\onlinecite{Girard87, Hutchinson90} and has been
employed by the Kernel Polynomial Method~\cite{Silver94,Silver96,Weisse06}.
Early related methods include Refs.~\onlinecite{Skilling89, Drabold93, Wang94}.
Since then, similar techniques have found practical applications in
quantum chemistry and electronic structure~\cite{Roeder97,Baer13,Neuhauser14,Gao15,Rabani15,Cytter18}.
As we discuss below, it is often preferable to design $R$ as a whole,
rather than work with its column vectors $r^{(s)}$ independently.

\subsection{Optimized probing}

Approximation~(\ref{eq:probing_Ne}) can be improved by optimizing
$R$ to take better advantage of the spatial decay properties of $f(H)_{ij}$.
The idea, presented in Refs.~\onlinecite{Bekas07,Tang12}, is to
construct $R$ such that the elements $(RR^{\dagger})_{ij}$ are zero
whenever $f(H)_{ji}$ is large, thus eliminating the largest contributions
to the error in Eq.~(\ref{eq:probing_err}). Here, we make use of
the physical property, to be discussed below, that $f(H)_{ij}$ decays
with spatial distance $r_{ij}$ between orbitals $i$ and $j$.

\begin{figure}
\includegraphics[width=0.8\columnwidth]{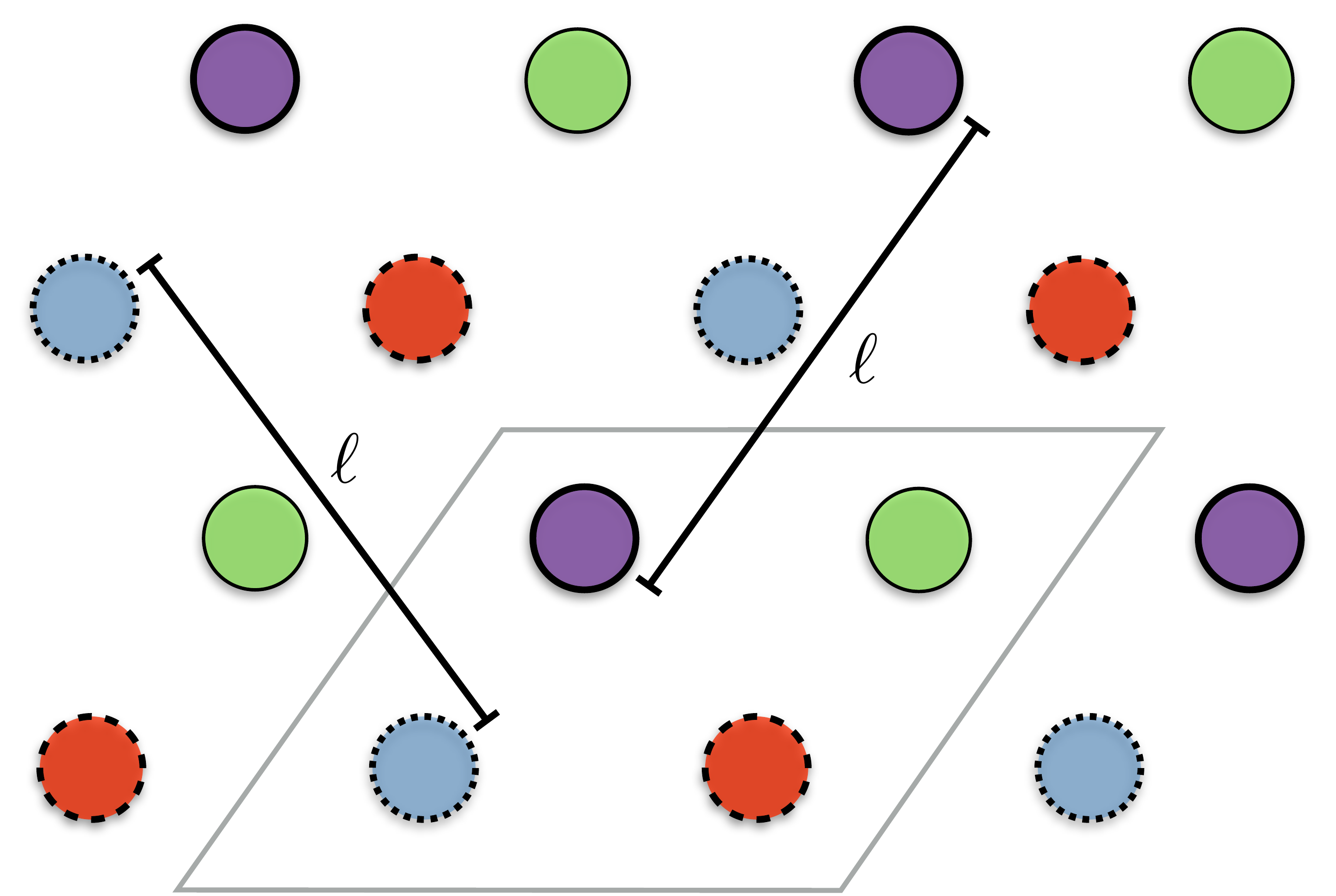}\caption{Coloring sites on the triangular lattice. With $S$ colors (here $4)$,
pairs of same-color sites are separated by at least $\ell=S^{1/2}$
lattice constants (here 2). In the general $d$-dimensional case we
expect to find a coloring such that same-color sites are separated
by a distance $\ell$ that scales like $S^{1/d}$.\label{fig:color}}
\end{figure}

The first step in designing the $N\times S$ matrix $R$ is to assign
a \emph{color} $c(i)\in\{1,2,\dots S\}$ to each localized orbital
$i$. We employ the heuristic that different colors should be assigned
to sites $i$ and $j$ whose separation $r_{ij}$ is small. That is,
given $S$ colors, we seek a coloring that satisfies
\begin{equation}
c(i)\neq c(j)\quad\mathrm{if}\quad r_{ij}<\ell,\label{eq:coloring_1}
\end{equation}
with the largest possible distance $\ell$. If the sites are distributed
roughly uniformly in $d$-dimensional space, we expect to find a coloring
that satisfies
\begin{equation}
\ell\sim S^{1/d}.\label{eq:coloring_2}
\end{equation}
An optimal strategy for coloring the triangular lattice is illustrated
in Fig.~\ref{fig:color}; observe that with $S$ colors, we can separate
same-color sites by $\ell=\sqrt{S}$ lattice constants. The case of
a one-dimensional lattice is even simpler: the optimal coloring, $c(i)=1+(i-1)\bmod S$,
separates same-color sites by $\ell=S$ lattice constants.

\begin{figure}
\includegraphics[width=1\columnwidth]{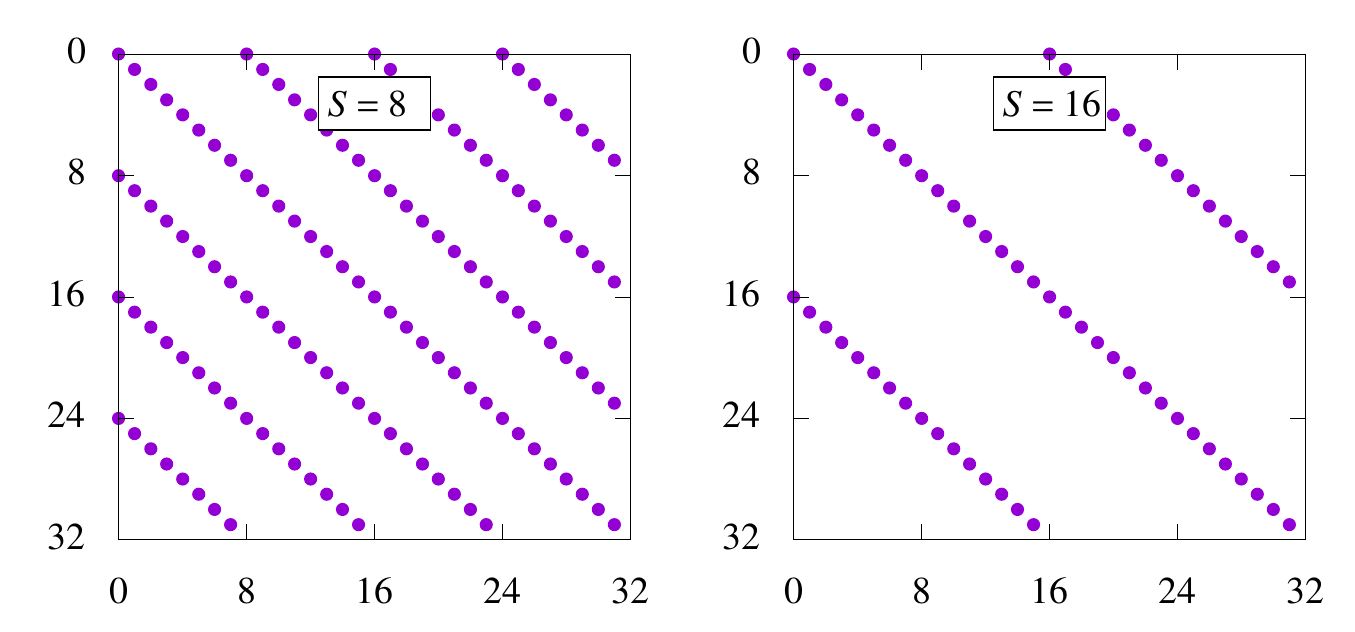}\caption{Structure of the matrix $RR^{\dagger}$ for an optimal coloring on
the one-dimensional lattice. The diagonal elements are exactly one.
The closest non-zero elements are a distance $S$ away from the diagonal
elements. This matrix structure enhances the convergence of trace
estimates $\protect\tr A\approx\protect\tr R^{\dagger}AR$ as a function
of the number of colors $S$.\label{fig:sparsity}}
\end{figure}

Given a coloring, we can replace the uncorrelated matrix of Eq.~(\ref{eq:R_uncorrelated})
with the optimized one
\begin{equation}
R_{is}=\delta_{c(i),s}\zeta_{i},\label{eq:R_correlated}
\end{equation}
where $\zeta_{i}$ are uncorrelated random numbers. The outer product
matrix becomes
\begin{equation}
(RR^{\dagger})_{ij}=\delta_{c(i),c(j)}\zeta_{i}\zeta_{j}^{\ast}.\label{eq:rrdag_correlated}
\end{equation}
As before, we constrain $|\zeta_{i}|=1$ such that $(RR^{\dagger})_{ii}=1$.
The off-diagonal elements $(RR^{\dagger})_{ij}$ are mostly zero,
except for orbital pairs $(i,j)$ that share the same color, $c(i)=c(j)$.
Figure~\ref{fig:sparsity} illustrates the sparsity structure of
$RR^{\dagger}$ in the one-dimensional case.

As before, the stochastic error is given by Eq.~(\ref{eq:probing_err}).
Its variance can be calculated by inserting Eq.~(\ref{eq:rrdag_correlated}).
After some analysis, we obtain a sum over same-color, off-diagonal
elements,
\begin{equation}
\mathrm{var}[\Delta N_{e}]=\sum_{i<j}\delta_{c(i),c(j)}|f(H)_{ij}|^{2}\quad\quad(R\,\mathrm{optimized)}.\label{eq:var_correlated}
\end{equation}
Compared to the uncorrelated result, Eq.~(\ref{eq:var_uncorrelated}),
we lose a prefactor of $1/S$ but gain the constraint $c(i)=c(j)$,
which eliminates all but $\sim1/S$ of the terms. The great advantage
of optimized probing is that the remaining terms correspond to orbital
pairs $(i,j)$ that satisfy $r_{ij}\geq\ell$, for which $f(H)_{ij}$
is small. To quantify the numerical advantage of probing, we must
first determine the actual smallness of relevant matrix elements $f(H)_{ij}$
and $g(H)_{ij}$.

Finally, we note that in the limit $S\rightarrow N$, each orbital
gets a unique color, $c(i)=i$, and the stochastic error in Eq.~(\ref{eq:var_correlated})
disappears. Our theoretical analysis will focus on the regime $1\ll S\ll N$.

\subsection{Spatial decay of density and energy matrices}

\begin{figure*}
\includegraphics[width=0.75\paperwidth]{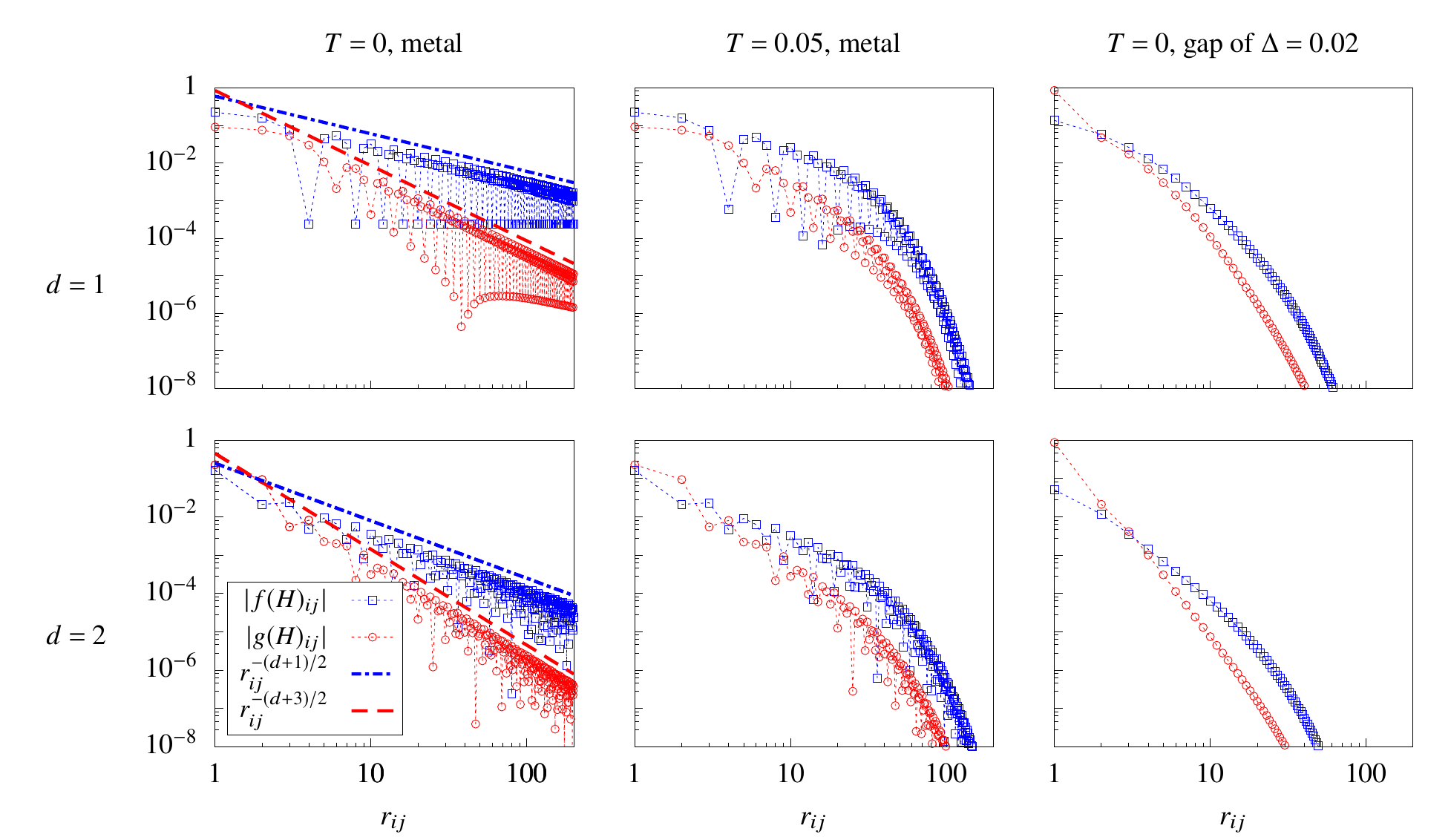}\caption{Decay of the density and energy matrices, $f(H)_{ij}$ and $g(H)_{ij}$,
as a function of distance $r_{ij}$ between localized orbitals $i$
and $j$. For zero-temperature metals (left column), we observe power
law scaling, $|f(H)_{ij}|\sim r_{ij}^{-(d+1)/2}$ and $|g(H)_{ij}|\sim r_{ij}^{-(d+3)/2}$,
where $d$ is the spatial dimension. If there is a finite temperature
$T$ (middle column), or band gap $\Delta$ (right column), the decay
becomes much faster. $T$ and $\Delta$ are measured in units of the
hopping constant for this model tight-binding system.\label{fig:matrix_decay}}
\end{figure*}
We focus our analysis on metallic systems at zero temperature, for
which the density matrix $f(H)$ decays most slowly, thus posing the
greatest numerical challenge. For simplicity, here we assume a single
electronic band with quadratic dispersion $\epsilon_{\mathbf{k}}=k^{2}/2$,
partially filled up to a chemical potential $\mu=k_{F}^{2}/2$. We
work in arbitrary spatial dimension $d$. At zero temperature the
Fermi function reduces to the Heaviside function, $f(\epsilon_{\mathbf{k}})=\Theta(\mu-\epsilon_{\mathbf{k}})$
and the real-space density matrix elements can be calculated by Fourier
transform~\cite{Vembu61,Barros13a}
\begin{align}
f(H)_{ij} & =(2\pi)^{-d}\int\d^{d}\mathbf{k}\,f(\epsilon_{\mathbf{k}})e^{-i\mathbf{k}\cdot\mathbf{r}_{ij}}\nonumber \\
 & =\left(\frac{k_{F}}{2\pi r_{ij}}\right)^{d/2}J_{d/2}(k_{F}r_{ij}).
\end{align}
We have assumed that the volume of the primitive cell is one.

For large argument $k_{F}r_{ij}$, the Bessel function of the first
kind scales asymptotically as
\begin{equation}
J_{d/2}(k_{F}r_{ij})\approx\sqrt{\frac{2}{\pi k_{F}r_{ij}}}\cos\left[k_{F}r_{ij}-\frac{\pi(d+1)}{4}\right].\label{eq:bessel_asympt}
\end{equation}
Ignoring oscillations, we conclude that the density matrix elements
decay as
\begin{equation}
|f(H)_{ij}|\sim f_{\mathrm{dec}}(r_{ij})=r_{ij}^{-(d+1)/2}.\label{eq:f_decay}
\end{equation}

We also consider the energy matrix, $g(H)$, defined via Eq.~(\ref{eq:en_func}).
At zero temperature, $g(\epsilon_{\mathbf{k}})=(\epsilon_{\mathbf{k}}-\mu)f(\epsilon_{\mathbf{k}})$
and the energy matrix elements become
\begin{align}
g(H)_{ij} & =(2\pi)^{-d}\int\d^{d}\mathbf{k}\,g(\epsilon_{\mathbf{k}})e^{-i\mathbf{k}\cdot\mathbf{r}_{ij}}\nonumber \\
 & =-\frac{k_{F}}{r_{ij}}\left(\frac{k_{F}}{2\pi r_{ij}}\right)^{d/2}J_{d/2+1}(k_{F}r_{ij}).
\end{align}
At large distances, $g(H)_{ij}$ decays as
\begin{equation}
|g(H)_{ij}|\sim g_{\mathrm{dec}}(r_{ij})=r_{ij}^{-(d+3)/2}.\label{eq:g_decay}
\end{equation}
We conclude that the energy matrix decays one power faster than the
density matrix, Eq.~(\ref{eq:f_decay}).

At small nonzero temperature, $T>0$, the decay of both $f_{\mathrm{dec}}(r_{ij})$
and $g_{\mathrm{dec}}(r_{ij})$ becomes exponential, $\exp(-cr_{ij}T))$,
for some constant $c$~\cite{Goedecker98}. Similarly, if the chemical
potential lies within a small band gap of width $\Delta$, the decay
also becomes exponential, $\exp(-cr_{ij}\Delta)$~\cite{Ismail-Beigi99}.
Once exponential decay is introduced, a new power law prefactor may
appear. For example, the density matrix for insulators with small
gap may actually scale as $|f(H)_{ij}|\sim r_{ij}^{-d/2}\exp(-cr_{ij}\Delta)$~\cite{He01,Taraskin02}.
These exponential decays are asymptotic upper bounds. For example,
a faster decay $\exp(-cr_{ij}\sqrt{\Delta})$ has been observed along
non-diagonal directions of a model insulator on the square lattice~\cite{Jedrzejewski04}.
Many of the above scaling bounds have been demonstrated with mathematical
rigor~\cite{Benzi13}.

Although Eqs.~(\ref{eq:f_decay}) and~(\ref{eq:g_decay}) were derived
in the context of a model isotropic material, the power law exponents
$(d+1)/2$ and $(d+3)/2$ are universal to bulk metals at zero temperature.
We demonstrate this numerically in the context of a simple tight-binding
model, $\mathcal{H}=t\sum_{\langle ij\rangle}c_{i}^{\dagger}c_{j}$,
with hoppings between nearest-neighbor sites, $\langle ij\rangle$.
We use dimensionless units for energy ($t=1$), temperature ($k_{B}=1$),
and length (lattice constant $a=1$). We study linear and square lattices,
$d=\{1,2\}$, with $N=\{10^{4},2000^{2}\}$ lattice sites, respectively.
For metals, we fix the electron number to quarter filling fraction.
To realize an insulating gap of width $\Delta$, we switch our model
system to half filling and introduce a uniform magnetic field to split
the spin-up and -down bands.

We use the Kernel Polynomial Method to expand $f(H)$ and $g(H)$,
as described in Appendix~\ref{sec:kpm}. To calculate the matrix
elements with high precision, we do not apply any stochastic approximation,
and we use an extremely large polynomial order, $M=10^{5}$.

Figure~\ref{fig:matrix_decay} shows the matrix decays for this model
system. At zero temperature, we observe the expected power laws of
$f_{\mathrm{dec}}(r_{ij})$ and $g_{\mathrm{dec}}(r_{ij})$. Scatter
is associated with the oscillatory nature of Bessel functions, Eq.~(\ref{eq:bessel_asympt}).
Introducing either finite temperature ($T=0.05$) or gap ($\Delta=0.02$)
leads to much faster matrix decay.

\subsection{Error analysis}

Armed with the decay properties of $f(H)$ and $g(H)$, we can now
quantify the stochastic errors $\Delta N_{e}$ and $\Delta\Omega$
for probing estimates of the electron number, $N_{e}=\tr f(H)$, and
grand canonical free energy, $\Omega=\tr g(H)$, respectively.

If the matrix $R$ is constructed as a concatenation of $S$ uncorrelated
random column vectors, then Eq.~(\ref{eq:var_uncorrelated}) gives
the variance of these errors as a double sum over orbitals. The first
sum, over $i$, is unconstrained, yielding a factor of system size
$N$. The second sum, over $j$, only contributes when orbitals $i$
and $j$ are local, due to the sufficiently fast spatial decay of
$f(H)$ and $g(H)$. We conclude
\begin{equation}
\mathrm{var}[\Delta N_{e}]\sim\mathrm{var}[\Delta\Omega]\sim N\times S^{-1}\quad\quad(R\,\mathrm{uncorrelated).}\label{eq:var_uncorrelated_result}
\end{equation}
The standard deviations of $\Delta N_{e}$ and $\Delta\Omega$ thus
scale like $\sqrt{N/S}$. Consequently, probing estimates of intensive
quantities such as $N_{e}/N$ and $\Omega/N$ actually improve with
increasing system size, which can be attributed to self-averaging~\cite{Silver94}.

We see a significant improvement when using the optimized matrix $R$
of Eq.~(\ref{eq:R_correlated}) with well selected colors $c(i)=\{1,2,\dots S\}$.
For metallic systems at zero temperature and spatial dimension $d$,
the variance of Eq.~(\ref{eq:var_correlated}) becomes
\begin{align}
\mathrm{var}[\Delta N_{e}] & \sim N\times S^{-(d+1)/d},\label{eq:var_optimized_ne_result}\\
\mathrm{var}[\Delta\Omega] & \sim N\times S^{-(d+3)/d}\quad\quad(R\,\mathrm{optimized)}.\label{eq:var_optimized_om_result}
\end{align}
The factor $N$ again appears due to a single unconstrained sum over
orbitals. The dependence on $S$ follows from the fact that same-color
sites are separated by a distance of at least $\ell\sim S^{1/d}$
{[}cf. Eqs.~(\ref{eq:coloring_1}) and~(\ref{eq:coloring_2}){]}.
Consequently, the largest terms in Eq.~(\ref{eq:var_correlated})
are of order $f_{\mathrm{dec}}(\ell){}^{2}$ and $g_{\mathrm{dec}}(\ell){}^{2}$,
whose scaling behaviors are given by Eqs.~(\ref{eq:f_decay}) and~(\ref{eq:g_decay}).
We note that $f(H)$ and $g(H)$ decay sufficiently fast such that
matrix elements $(i,j)$ with separation $r_{ij}\gg\ell$ do not contribute
significantly to the above variances.

\begin{figure}
\includegraphics[width=1\columnwidth]{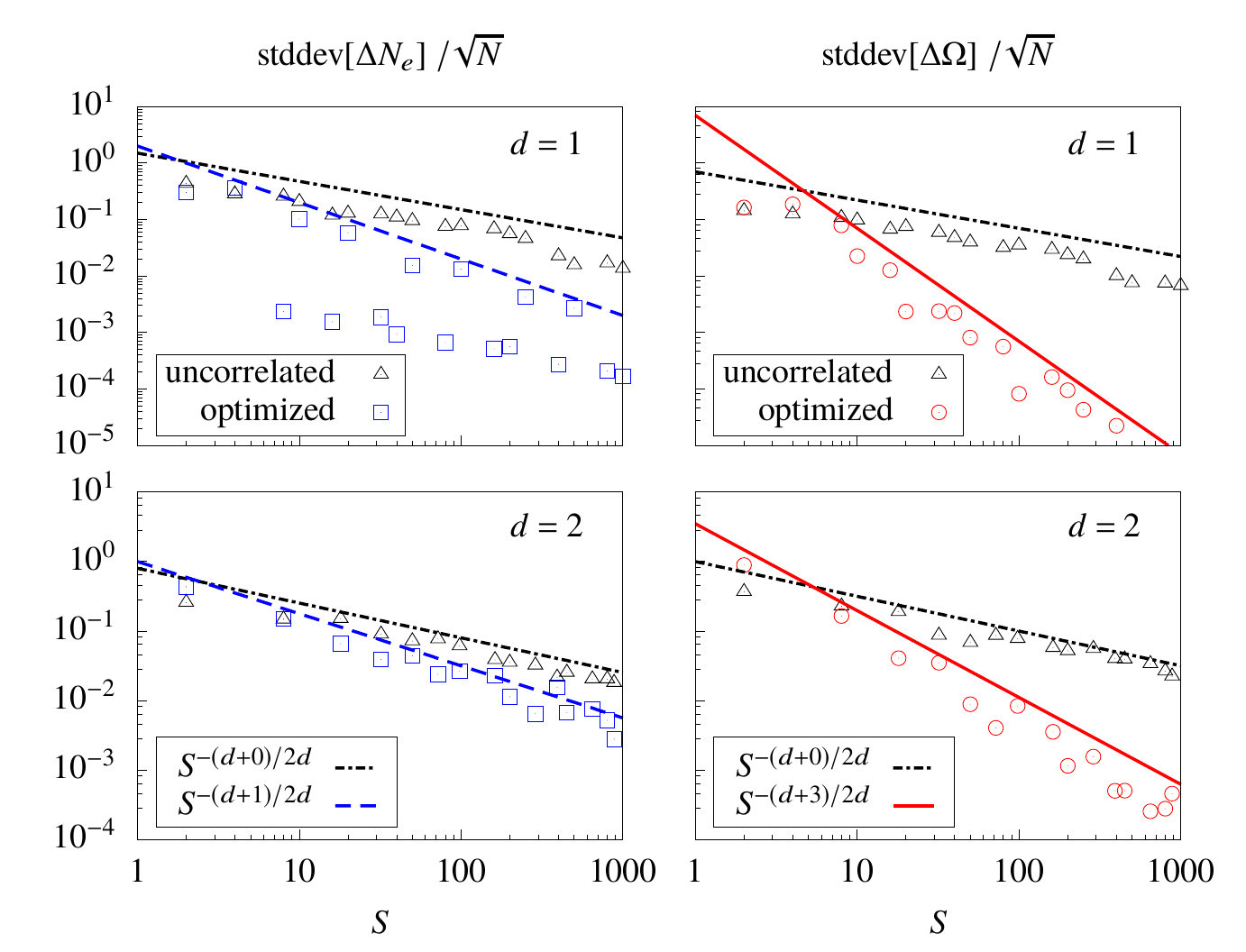}\caption{Stochastic errors for probing estimates of the electron number $N_{e}$
and free energy $\Omega$. With $S$ uncorrelated random vectors,
Eq.~(\ref{eq:R_uncorrelated}), the errors scale as $S^{-1/2}$.
Optimized probing, Eq.~(\ref{eq:R_correlated}), yields much smaller
errors. For a zero-temperature metallic system, $\Delta N_{e}\sim S^{-(d+1)/2d}$
and $\Delta\Omega\sim S^{-(d+3)/2d}$. \label{fig:trace_estimates}}
\end{figure}

Figure~\ref{fig:trace_estimates} shows numerical estimation of the
probing errors for zero-temperature metals. Again, we use a nearest
neighbor tight-binding model in $d=\{1,2\}$ dimensions, and quarter
electron filling. The lattice sizes are $N=\{10^{4},1260^{2}\}$ respectively.
We perform the numerics with Chebyshev polynomial order $M=3000$.
To estimate the standard deviations of probing errors, we repeat each
probing calculation 10 times using independent samples of the random
matrix $R$. Due to translation invariance of the model, we can extract
an independent estimate of the stochastic errors at each lattice site,
over which we average. The resulting estimates of $\mathrm{stddev}[\Delta N_{e}]$
and $\mathrm{stddev}[\Delta\Omega]$ verify the power law scaling
predicted by Eqs.~(\ref{eq:var_optimized_ne_result}) and~(\ref{eq:var_optimized_om_result}).
The scatter, as a function of $S$, is not a sampling artifact; it
arises due to the oscillatory decay of $f(H)_{ij}$ and $g(H)_{ij}$.

\section{Density matrix estimation\label{sec:density_matrix_estimation}}

\subsection{Direct probing}

The trace approximation of Eq.~(\ref{eq:probing_Ne}) generalizes
to an approximation for individual density matrix elements,\footnote{In actual practice, we use a symmetrized probing approximation for
the density matrix: $f(H)\approx[f(H)RR^{\dagger}+RR^{\dagger}f(H)]/2$.} 
\begin{equation}
f(H)\approx f(H)RR^{\dagger}.\label{eq:f_probing}
\end{equation}
Taking the trace of both sides recovers Eq.~(\ref{eq:probing_Ne})
exactly. In a numerical implementation, we do not construct the full
matrix $RR^{\dagger}$ explicitly. Instead, we first build $f(H)R$
and then use it to calculate only the desired elements $f(H)_{ij}$.
Details are discussed below in Sec.~\ref{subsec:numerics}.

The stochastic error of the direct approximation is
\begin{equation}
\Delta f(H)^{\mathrm{direct}}=f(H)(RR^{\dagger}-I).\label{eq:err_direct}
\end{equation}
Repeating the analysis of the previous section, we find that its variance
scales as
\begin{equation}
\mathrm{var}[\Delta f(H)_{ij}^{\mathrm{direct}}]\sim\begin{cases}
S^{-1} & (R\,\mathrm{uncorrelated)}\\
S^{-(d+1)/d} & (R\,\mathrm{optimized)}
\end{cases},\label{eq:var_f_direct}
\end{equation}
for the $R$ matrices specified in Eqs.~(\ref{eq:R_uncorrelated})
and~(\ref{eq:R_correlated}), respectively. To see that these results
are consistent with Eqs.~(\ref{eq:var_uncorrelated_result}) and~(\ref{eq:var_optimized_ne_result}),
we first observe that $N_{e}=\sum_{i}f(H)_{ii}$. Consequently, $\mathrm{var}[\Delta N_{e}]$
decomposes into a sum over $N$ contributions $\sum_{i}\mathrm{var}[\Delta f(H)_{ii}^{\mathrm{direct}}]$,
and thus scales like $N\times\mathrm{var}[\Delta f(H)_{ij}^{\mathrm{direct}}]$.
An overall factor of system size $N$ thus appears in estimates of
\emph{extensive }quantities such as $N_{e}$ and $\Omega$, but not
local quantities such as $f(H)_{ij}$. Note that $\mathrm{var}[\Delta f(H)_{ij}^{\mathrm{direct}}]$
is roughly independent of the choice of orbitals $i$ and $j$ provided
that their distance $r_{ij}$ is small, which we will assume.

\subsection{Gradient-based probing}

A key observation in this paper is that it is possible to achieve
faster stochastic convergence than with direct probing. Inserting
Eq.~(\ref{eq:density_matrix}) into~(\ref{eq:free_energy}), and
applying approximation~(\ref{eq:probing_omega}), we find,
\begin{equation}
f(H)=\frac{\d}{\d H^{T}}\tr g(H)\approx\frac{\d}{\d H^{T}}\tr g(H)RR^{\dagger},\label{eq:f_autodiff}
\end{equation}
with error
\begin{equation}
\Delta f(H)^{\mathrm{grad}}=\frac{\d}{\d H^{T}}\tr g(H)(RR^{\dagger}-I).\label{eq:err_autodiff}
\end{equation}
Interestingly, the stochastic errors of direct~(\ref{eq:f_probing})
and gradient-based~(\ref{eq:f_autodiff}) approximation schemes are
\emph{not }the same,
\[
\Delta f(H)^{\mathrm{direct}}\neq\Delta f(H)^{\mathrm{grad}}.
\]
To demonstrate the inequality in a simple context, consider substitutions
$f\mapsto nH^{n-1}$ and $g\mapsto H^{n}$ for integer $n$. Then
\begin{equation}
nH^{n-1}=\frac{\d}{\d H^{T}}\tr H^{n},
\end{equation}
but
\begin{align}
nH^{n-1}RR^{\dagger} & \neq\nonumber \\
\sum_{m=1}^{n} & H^{m-1}RR^{\dagger}H^{n-m}=\frac{\d}{\d H^{T}}\tr H^{n}RR^{\dagger}.
\end{align}
Inequality stems from the fact that $H$ and $RR^{\dagger}$ do not
commute. Observe that the gradient-based approximation (i.e., the
right hand side) may benefit from cancellations between $n$ different
approximations, each unbiased. %
The density matrix $f(H)$ may be expanded in powers of $H$, suggesting
that $\Delta f(H)^{\mathrm{grad}}$ may similarly be smaller than
$\Delta f(H)^{\mathrm{direct}}$. After careful analysis (Appendix~\ref{sec:gradient_decay})
we find, 
\begin{equation}
\mathrm{var}[\Delta f(H)_{ij}^{\mathrm{grad}}]\sim S^{-(d+2)/d}\quad\quad(R\,\mathrm{optimized)},\label{eq:var_f_deriv}
\end{equation}
which is indeed superior to the direct probing approach, Eq.~(\ref{eq:var_f_direct}).
Intuitively, we associate the smaller error of gradient-based probing
with the faster decay of the energy matrix, relative to the density
matrix. Surprisingly, the exponent $(d+2)/d$ is new, and halfway
between the exponents associated with the decay of $f(H)$ and $g(H)$
{[}cf. Eqs.~(\ref{eq:var_optimized_ne_result}) and~(\ref{eq:var_optimized_om_result}){]}.

\subsection{\label{subsec:numerics}Numerics}

\begin{figure}
\includegraphics[width=1\columnwidth]{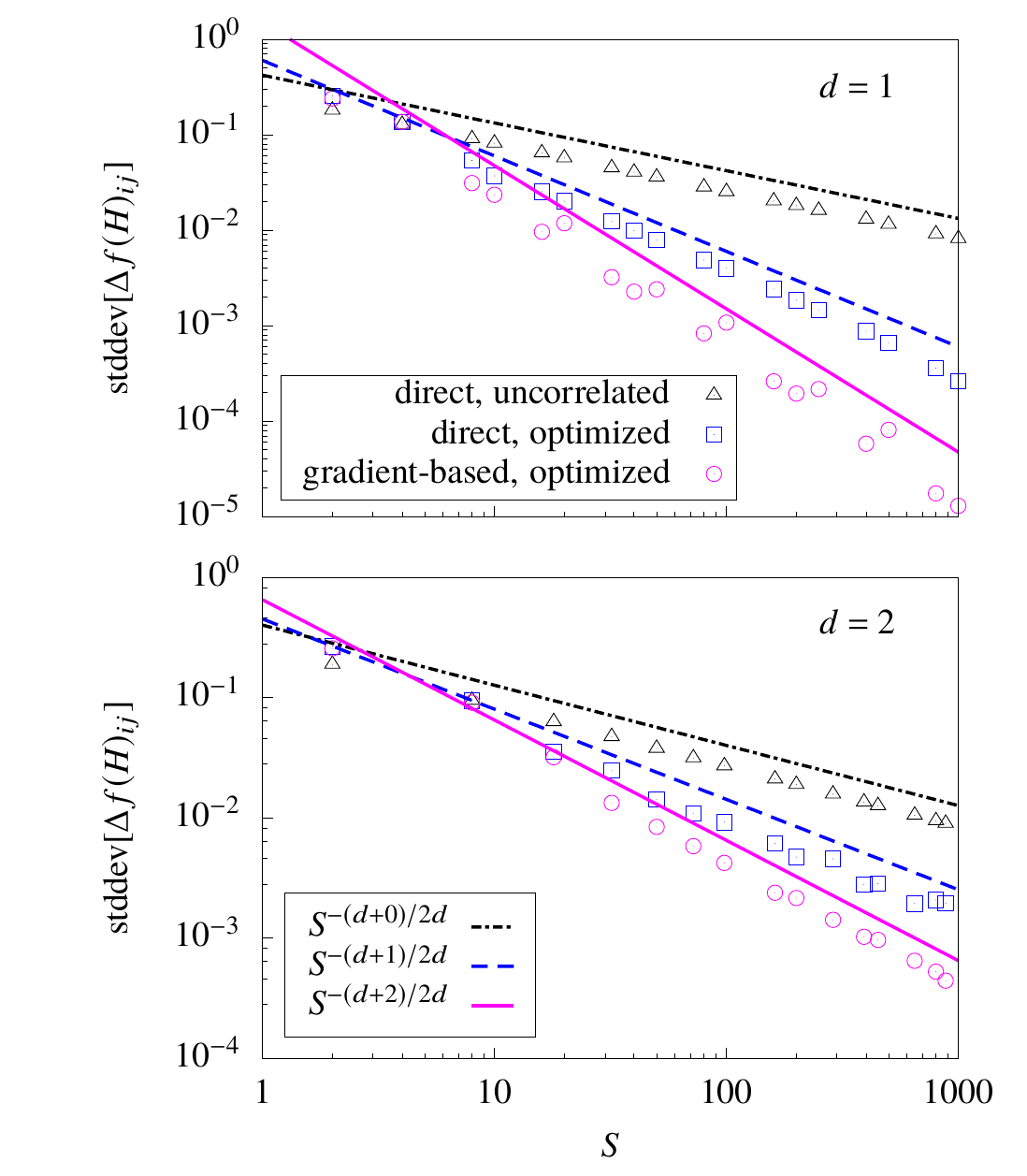}\caption{Stochastic errors for probing estimates of local density matrix elements
$f(H)_{ij}$. We consider a zero-temperature metallic system. Using
direct probing, $f(H)\approx f(H)RR^{\dagger}$, with $S$ uncorrelated
random vectors, Eq.~(\ref{eq:R_uncorrelated}), the stochastic error
scales like $\Delta f\sim S^{-1/2}$. When $R$ is optimized according
to Eq.~(\ref{eq:R_correlated}), the error scales like, $\Delta f\sim S^{-(d+1)/2d}$.
Our new gradient-based approximation, $f(H)\approx(\partial/\partial H^{T})\,\protect\tr g(H)RR^{\dagger}$,
achieves the smallest error: $\Delta f\sim S^{-(d+2)/2d}$.\label{fig:autodiff}}
\end{figure}
Figure~\ref{fig:autodiff} illustrates the accuracy of various approximation
schemes for estimating $f(H)_{ij}$. We use the same tight-binding
model and methods as in Fig.~\ref{fig:trace_estimates}. Here, however,
we measure the standard deviation of error $\Delta f(H)_{ij}$ for
probing estimates of individual matrix elements $f(H)_{ij}$. We take
$i$ and $j$ to be nearest-neighbor lattice sites $\langle ij\rangle$,
but the same asymptotic scaling holds for any local matrix element
(next nearest-neighbors, etc.). We confirm the power laws predicted
in Eqs.~(\ref{eq:var_f_direct}) and~(\ref{eq:var_f_deriv}) for
zero-temperature metals.

The details of our numerical implementation are as follows. For direct
probing, Eq.~(\ref{eq:f_probing}), we use the methods presented
in Appendix~\ref{sec:kpm}. We start with the Chebyshev polynomial
expansion $f(H)\approx f_{M}(H)=\sum_{m=0}^{M-1}c_{m}^{(f)}T_{m}(H)$
of Eq.~(\ref{eq:phi_approx}). Next, we evaluate the matrix product
$f_{M}(H)R$ as a linear combination of matrices $\alpha_{m}=T_{m}(H)R$,
which are calculated recursively using Eq.~(\ref{eq:alpha}). Finally,
we take the outer product, $f(H)_{ij}\approx\sum_{s=1}^{S}[f_{M}(H)R]_{is}R_{js}^{*}$,
for desired elements $(i,j)$. In typical applications, we require
$f(H)_{ij}$ only if $H_{ij}$ is non-vanishing, i.e., if the distance
$r_{ij}$ between orbitals $(i,j)$ is very small. The total computational
cost to approximate $f(H)$ thus scales like $\mathcal{O}(NMS)$.

The gradient-based probing approximation of Eq.~(\ref{eq:f_autodiff})
is more subtle to implement. We begin with the free energy approximation
described in Appendix~\ref{sec:kpm}. Specifically, we use the recursive
procedure defined by Eqs.~(\ref{eq:mu_approx})–(\ref{eq:omega_approx_stoch})
to calculate $\Omega\approx\tilde{\Omega}=\tr R^{\dagger}f_{M}(H)R$.
Taking the exact derivative of the approximate free energy yields
the desired density matrix approximation, $f(H)\approx\d\tilde{\Omega}/\d H^{T}$.
Appendix~\ref{sec:autodiff} describes the procedure to calculate
matrix elements $\d\tilde{\Omega}/\d H_{ji}$ using reverse-mode automatic
differentiation. Crucially, we calculate all relevant matrix elements
simultaneously, such that the procedures to estimate $\Omega$ and
$f(H)$ both scale like $\mathcal{O}(NMS)$.

We save a factor of 2 in the computational cost by using a product
identity for Chebyshev polynomials, as described in Appendix~\ref{sec:recipes}.

\section{Conclusions}

Our aim is efficient estimation of the density matrix $f(H)$, where
$H$ is the single-particle Hamiltonian. The greatest numerical challenge
appears for metals at zero temperature; in this case, $f(H)_{ij}$
decays like $r_{ij}^{-(d+1)/2}$, where $r_{ij}$ is the distance
between orbitals and $d$ the spatial dimension.

In a direct probing approach, one may approximate
\begin{equation}
f(H)\approx f(H)RR^{\dagger},
\end{equation}
where $R$ is a suitable $N\times S$ matrix. If the elements of $R$
are random and uncorrelated, the stochastic error associated with
estimates of local density matrix elements scales as $\Delta f\sim S^{-1/2}$.
Better approaches take advantage of the spatial decay of $f(H)$.
Optimized probing carefully assigns a color $c(i)\in\{1,2,\dots S\}$
to each local orbital $i$, such that nearby orbitals have different
colors~\cite{Tang12}. Then the $R$ matrix of Eq.~(\ref{eq:R_correlated})
yields improved approximations, with error $\Delta f\sim S^{-(d+1)/2d}$
for metals at zero temperature.

In this paper, we introduce a new \emph{gradient-based} probing technique,
\begin{equation}
f(H)\approx\frac{\d}{\d H^{T}}\tr g(H)RR^{\dagger},\label{eq:derivative_based_concl}
\end{equation}
where $\d g(x)/\d x=f(x)$. This approximation would become exact
if we were to replace $RR^{\dagger}$ with the identity. We show that
the energy matrix elements $g(H)_{ij}$ decay like $r_{ij}^{-(d+3)/2}$
for metals at zero temperature. Equation~(\ref{eq:derivative_based_concl})
with optimized $R$ leverages this faster matrix decay; careful analysis
shows that the stochastic error scales like $\Delta f\sim S^{-(d+2)/2d}$,
which we have confirmed numerically.

By applying reverse-mode automatic differentiation to the Kernel Polynomial
Method, we demonstrate an efficient implementation strategy for gradient-based
probing. The computational cost to estimate $\mathcal{O}(N)$ local
elements $f(H)_{ij}$ scales like $\mathcal{O}(NMS)$ where $M$ is
the polynomial expansion order.

Previous linear-scaling methods have largely focused on systems for
which the density matrix decays exponentially with distance. In such
cases, gradient-based probing also converges exponentially quickly.
Quantitative comparison with previous state-of-the-art implementations~\cite{Cawkwell12,VandeVondele12,Lin14}
will require experimentation, and is a topic for future work. A clear
advantage of gradient-based probing, however, is that it continues
to provide a high-quality, \emph{linear-scaling, }and\emph{ unbiased}
approximation to density matrix elements for metals at very low temperatures.
Our GPU-optimized implementation has enabled simulations of magnetic
moment dynamics on lattices of unprecedented size~\cite{Barros13,Barros14b,Ozawa16,Wang16,Ozawa17,Ozawa17a,Chern18}.
Gradient-based probing can potentially also be applied to realistic
quantum chemistry models, e.g. Kohn-Sham density functional theory
along the lines of Refs.~\onlinecite{Baer13, Cytter18}.

\section*{Supplementary Material}

See supplementary material for a stand-alone, minimal \textsf{Python}
code that demonstrates gradient-based probing.
\begin{acknowledgments}
We thank A. M. N. Niklasson, A. F. Voter, H. Suwa, and the anonymous
referees for their encouragement and helpful suggestions. Work performed
at LANL was supported by the Laboratory Directed Research and Development
(LDRD) program. C. D. B. and G.-W. C. were supported by the Center
for Materials Theory as a part of the Computational Materials Science
(CMS) program, funded by the U.S. Department of Energy, Office of
Science, Basic Energy Sciences, Materials Sciences and Engineering
Division.
\end{acknowledgments}

\appendix 

\section{\label{sec:kpm}Kernel Polynomial Method}

\subsection{Expansion of the density of states}

The density of states,
\begin{equation}
\rho(x)=\sum_{\epsilon}\delta(x-\epsilon),\label{eq:dos}
\end{equation}
is a representation of the eigenvalues $\epsilon$ of the Hamiltonian
$H$. The Kernel Polynomial Method~\cite{Silver94,Silver96,Weisse06}
approximates
\begin{equation}
\rho(x)\approx\rho_{M}(x)=\sum_{m=0}^{M-1}g_{m}^{M}\frac{w(x)}{q_{m}}\mu_{m}T_{m}(x)\label{eq:rho_M}
\end{equation}
using Chebyshev polynomials $T_{m}(x)=\cos(m\arccos x)$ up to order
$M$, and is valid in the range $|x|\leq1$. The moments

\begin{equation}
\mu_{m}=\int_{-1}^{+1}\rho(x)T_{m}(x)\d x=\tr T_{m}(H)\label{eq:mu1}
\end{equation}
are essentially the Fourier transform of $\rho(x)$ in the variable
$\theta=\arccos x$. The trace representation is valid assuming that
the eigenvalues of $H$ satisfy $|\epsilon|<1$. Given an unscaled
Hamiltonian $H_{0}$, we use the Lanczos method to estimate its extreme
eigenvalues $\epsilon_{\min}$ and $\epsilon_{\max}$~\cite{Lanczos50},
from which we define $H=2(H_{0}-\epsilon_{\mathrm{min}})/(\epsilon_{\mathrm{max}}-\epsilon_{\mathrm{min}})-I$.

Equation~(\ref{eq:rho_M}) with $M\rightarrow\infty$ and $g_{m}^{M}\rightarrow1$
follows from completeness of the Chebyshev polynomials and orthogonality
under the weights
\begin{align}
w(x) & =(1-x^{2})^{-1/2},\\
q_{m} & =\begin{cases}
\pi & \quad m=0\\
\pi/2 & \quad m\geq1
\end{cases}.\label{eq:q_m}
\end{align}
\begin{figure}
\includegraphics{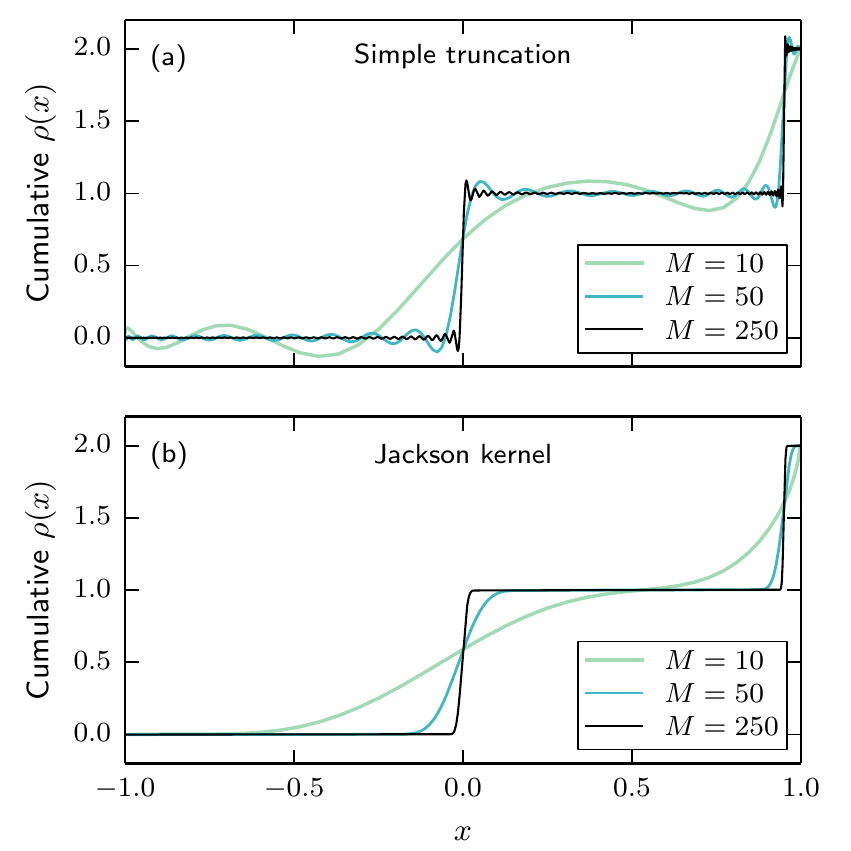}\caption{\label{fig:increasingM}Kernel polynomial approximation, Eq.~(\ref{eq:rho_M}),
to the density of states $\rho_{M}(x)\approx\rho(x)=\delta(x)+\delta(x-0.9)$.
To aid visualization, we plot $\int_{-1}^{x}\rho_{M}(x')\,\protect\d x'$.
\textbf{(a)} Simple truncation of the Chebyshev expansion ($g_{m}^{M}=1$)
leads to Gibbs oscillations. \textbf{(b)} The coefficients~(\ref{eq:jackson})
associated with the Jackson kernel optimally damp such artifacts.
Note that energy resolution is best when $x\rightarrow\pm1$.}
\end{figure}

Simple truncation, $g_{m}^{M}=1$, at finite $M$ would lead to Gibbs
oscillations. To damp these oscillations, we instead select coefficients
\begin{equation}
g_{m}^{M}=\frac{\left(M-m+1\right)\cos\frac{\pi m}{M+1}+\sin\frac{\pi m}{M+1}\cot\frac{\pi}{M+1}}{M+1}\label{eq:jackson}
\end{equation}
corresponding to the Jackson kernel~\cite{Jackson11,Jackson12,Weisse06}.
With this choice, $\rho_{M}(x)$ is a strictly non-negative approximation
to $\rho(x)$, and converges uniformly in the limit $M\rightarrow\infty$.
Figure~\ref{fig:increasingM} illustrates $\rho_{M}(x)$ for various
$M$.

\subsection{Expansion of matrix functions}

The approximate density of states $\rho_{M}(x)$ enables trace estimates,
\begin{align}
\tr\phi(H)= & \int_{-1}^{+1}\rho(x)\phi(x)\,\d x\nonumber \\
 & \approx\int_{-1}^{+1}\rho_{M}(x)\phi(x)\,\d x=\sum_{m=0}^{M-1}c_{m}^{(\phi)}\mu_{m},\label{eq:tr_phi_approx}
\end{align}
for any matrix function $\phi(H)$. Estimates of the free energy,
$\Omega=\tr g(H)$, and electron number, $N_{e}=\tr f(H)$, follow
directly. The coefficients
\begin{equation}
c_{m}^{(\phi)}=\int_{-1}^{+1}g_{m}^{M}\frac{w(x)}{q_{m}}\phi(x)T_{m}(x)\d x\label{eq:c_m}
\end{equation}
may be interpreted as a Chebyshev polynomial expansion of $\phi(H)$,
\begin{equation}
\phi(H)\approx\phi_{M}(H)=\sum_{m=0}^{M-1}c_{m}^{(\phi)}T_{m}(H).\label{eq:phi_approx}
\end{equation}
Direct evaluation of $\tr\phi_{M}(H)$ using Eq.~(\ref{eq:mu1})
reproduces the same approximation as in Eq.~(\ref{eq:tr_phi_approx}).

The definite integrals of Eq.~(\ref{eq:c_m}) require care to evaluate.
Chebyshev-Gauss quadrature~\cite{Abramowitz72} gracefully handles
the $x=\pm1$ singularities of the weight function $w(x)=1/\sqrt{1-x^{2}}$.
The result is

\begin{align}
c_{m}^{(\phi)} & \approx\frac{\pi g_{m}^{M}}{N_{M}q_{m}}\sum_{n=0}^{N_{M}-1}\cos(m\theta_{n})\phi(\cos\theta_{n}),
\end{align}
with $\theta_{n}=\pi(n+\frac{1}{2})/N_{M}$. A reasonable choice for
the number of quadrature points is $N_{M}=2M$, where $M$ is the
polynomial expansion order~\cite{Weisse06}. The fast discrete cosine
transform of the second kind (DCT-II) can be used to calculate all
$c_{m}^{(\phi)}$ at cost $\mathcal{O}(N_{M}\ln N_{M})$.

\subsection{Stochastic approximation}

The utility of the Kernel Polynomial Method is that the Chebyshev
moments $\mu_{m}$ may be directly estimated. The Chebyshev polynomials
satisfy the recurrence relation,
\begin{equation}
T_{m}\left(H\right)=\left\{ \begin{array}{ll}
1 & \quad m=0\\
H & \quad m=1\\
2HT_{m-1}(H)-T_{m-2}(H) & \quad m\geq2
\end{array}.\right.\label{eq:cheby_recurs}
\end{equation}
Rather than calculate $\mu_{m}=\tr T_{m}(H)$ directly, we apply the
probing approximation of Eqs.~(\ref{eq:probing_Ne}) and~(\ref{eq:probing_omega}),
\begin{equation}
\mu_{m}\approx\tilde{\mu}_{m}=\tr R^{\dagger}T_{m}(H)R=\tr R^{\dagger}\alpha_{m},\label{eq:mu_approx}
\end{equation}
where $R$ is an $N\times S$ random matrix, e.g., as defined in Eq.~(\ref{eq:R_uncorrelated})
or~(\ref{eq:R_correlated}). The $N\times S$ matrices $\alpha_{m}\equiv T_{m}(H)R$
may be calculated recursively,

\begin{equation}
\alpha_{m}=\left\{ \begin{array}{ll}
R & \quad m=0\\
HR & \quad m=1\\
2H\alpha_{m-1}-\alpha_{m-2} & \quad m\geq2
\end{array}.\right.\label{eq:alpha}
\end{equation}
If we assume sparse $H$ with $\mathcal{O}(N)$ nonzero elements,
each matrix-matrix multiplication requires $\mathcal{O}(NS)$ operations,
and the cost to estimate all moments scales like $\mathcal{O}(NMS)$.

Substituting the approximate moments $\tilde{\mu}_{m}$ into Eq.~(\ref{eq:rho_M})
yields an approximate density of states. The corresponding free energy
and electron number approximations follow from Eq.~(\ref{eq:tr_phi_approx}),
\begin{align}
\Omega & \approx\tilde{\Omega}=\sum_{m=0}^{M-1}c_{m}^{(g)}\tilde{\mu}_{m},\label{eq:omega_approx_stoch}\\
N_{e} & \approx\tilde{N}_{e}=\sum_{m=0}^{M-1}c_{m}^{(f)}\tilde{\mu}_{m},\label{eq:ne_approx_stoch}
\end{align}
with coefficients $c_{m}^{(\phi)}$ defined by Eq.~(\ref{eq:c_m}). 

\section{\label{sec:autodiff}Automatic differentiation of stochastic trace
estimates}

\begin{figure}[tb]
\centering

\includegraphics[width=0.95\columnwidth]{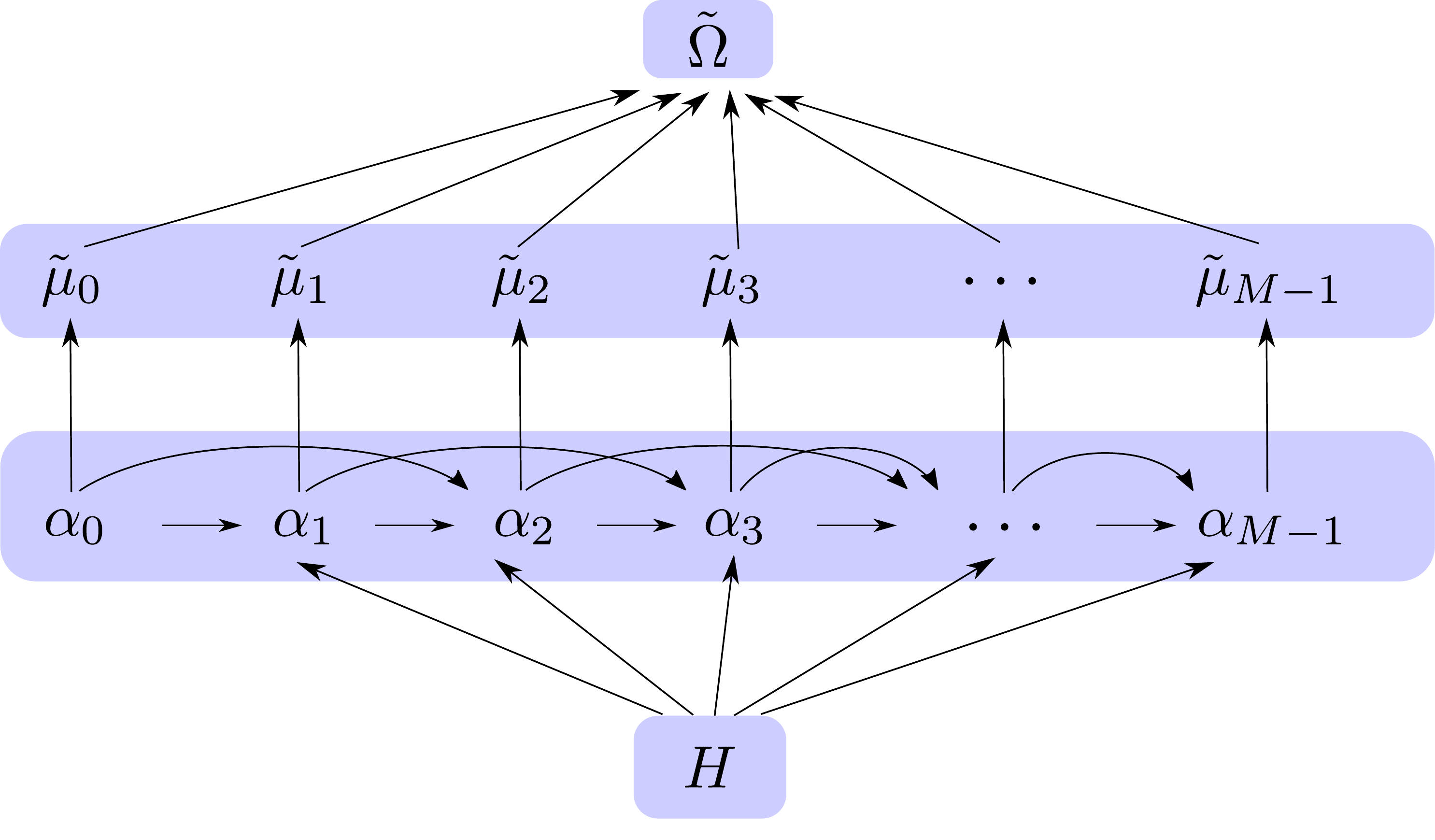}\caption{A graphical representation of the linear-scaling procedure to approximate
the free energy $\tilde{\Omega}(H)$. This directed computational
graph makes explicit all functional dependencies between intermediate
terms, $H\rightarrow\alpha_{m}\rightarrow\tilde{\mu}_{m}\rightarrow\tilde{\Omega}$.
Reverse-mode automatic differentiation calculates the derivative $\protect\d\tilde{\Omega}/\protect\d H^{T}$
for all relevant matrix elements by traversing this graph backwards. }
\label{fig:autodiff_tree}
\end{figure}

Equations~(\ref{eq:mu_approx})–(\ref{eq:omega_approx_stoch}) are
the basis for a linear-scaling numerical procedure to calculate the
approximate free energy $\tilde{\Omega}$. The computational graph
representing this recursive procedure is illustrated in Fig.~\ref{fig:autodiff_tree}.

Using the technique of reverse-mode automatic differentiation~\cite{Griewank89,Barros13},
we calculate the derivative $\d\tilde{\Omega}/\d H^{T}$ for many
matrix elements simultaneously by traversing the graph backward. First,
a remark on notation: We use partial derivatives $\partial z_{j}(z_{i},\dots)/\partial z_{i}$
in reference to the direct functional dependence of $z_{j}$ on $z_{i}$.
That is, each partial derivative corresponds to a single arrow in
the computational graph. In contrast, we use total derivatives $\d\tilde{\Omega}/\d z_{i}$
to denote the \emph{complete} transitive dependence of $\tilde{\Omega}$
on $z_{i}$. The chain rule states that $\d\tilde{\Omega}/\d H=p_{1}+p_{2}+\dots$
is a sum over all paths from $H$ to $\tilde{\Omega}$ in the graph.
Each path is a chained product of partial derivatives, e.g., $p=(\partial\tilde{\Omega}/\partial z_{n})(\partial z_{n}/\partial z_{n-1})\dots(\partial z_{1}/\partial H)$
where $H\rightarrow z_{1}\rightarrow z_{2}\rightarrow\dots\,\tilde{\Omega}$
are connected by arrows in Fig.~\ref{fig:autodiff_tree}.

Reverse-mode automatic differentiation works by expressing the desired
sum-over-paths $\d\tilde{\Omega}/\d H$ using terms $\d\tilde{\Omega}/\d z_{j}$
associated with intermediate paths, $z_{j}\rightarrow z_{j+1}\rightarrow\dots\,\tilde{\Omega}$.
We recursively move the starting point $z_{j}$ backwards along the
computational graph, constructing ever longer sums-over-paths, until
we finally reach the input matrix $H$. Crucially, reverse-mode automatic
differentiation visits each intermediate variable $z_{j}$ only \emph{once},
yet is able to produce all desired matrix elements of $\d\tilde{\Omega}/\d H$.

To explain the recursive procedure, we first consider a simple example.
Suppose we have already calculated $\d\tilde{\Omega}/\d z_{3}$ and
$\d\tilde{\Omega}/\d z_{2}$. Suppose further that $z_{1}$ is an
intermediate variable that appears only in the functional dependencies
for $z_{3}(z_{1},\dots)$ and $z_{2}(z_{1},\dots)$. Then the chain
rule states $\d\tilde{\Omega}/\d z_{1}=(\d\tilde{\Omega}/\d z_{3})(\partial z_{3}/\partial z_{1})+(\d\tilde{\Omega}/\d z_{2})(\partial z_{2}/\partial z_{1})$.
In other words, knowing $\d\tilde{\Omega}/\d z_{3}$ and $\d\tilde{\Omega}/\d z_{2}$
allows us to calculate $\d\tilde{\Omega}/\d z_{1}$. Note that the
partial derivatives $\partial z_{3}/\partial z_{1}$ and $\partial z_{2}/\partial z_{1}$
are never required by the procedure again. We keep working backwards
in this way until we eventually reach $\d\tilde{\Omega}/\d H$.

Now we derive the full recursive automatic differentiation procedure.
Partial derivatives (i.e., the arrows in Fig.~\ref{fig:autodiff_tree})
follow immediately from Eq.~(\ref{eq:omega_approx_stoch}),

\begin{equation}
\frac{\partial\tilde{\Omega}}{\partial\tilde{\mu}_{m}}=c_{m}^{(g)},
\end{equation}
from Eq.~(\ref{eq:mu_approx}),
\begin{equation}
\frac{\partial\tilde{\mu}_{m}}{\partial\alpha_{m;lk}}=R_{lk}^{\ast},
\end{equation}
and from Eq.~(\ref{eq:alpha}),

\begin{align}
\frac{\partial\alpha_{m+1;lk}}{\partial\alpha_{m;ij}} & =2H_{li}\delta_{kj},\\
\frac{\partial\alpha_{m+2;lk}}{\partial\alpha_{m;ij}} & =-\delta_{li}\delta_{kj},\\
\frac{\partial\alpha_{m;lk}}{\partial H_{ij}} & =\begin{cases}
\delta_{li}\alpha_{0;jk} & \quad m=1\\
2\delta_{li}\alpha_{m-1;jk} & \quad m>1
\end{cases}.
\end{align}
Our procedure to calculate $\d\tilde{\Omega}/\d H^{T}$ will require
first constructing $\d\tilde{\Omega}/\d\tilde{\mu}_{m}$ and then
$\d\tilde{\Omega}/\d\alpha_{m}$. Beginning at the top of Fig.~\ref{fig:autodiff_tree},
there is only one path from each $\tilde{\mu}_{m}$ to $\tilde{\Omega}$,
so

\begin{equation}
\frac{\d\tilde{\Omega}}{\d\tilde{\mu}_{m}}=c_{m}^{(g)}.
\end{equation}
Next, we seek $\d\tilde{\Omega}/\d\alpha_{m}$, which can be factorized
using recursively calculated quantities,
\begin{align}
\frac{\d\tilde{\Omega}}{\d\alpha_{m}} & =\frac{\d\tilde{\Omega}}{\d\tilde{\mu}_{m}}\frac{\partial\tilde{\mu}_{m}}{\partial\alpha_{m}}+\sum_{lk}\frac{\d\tilde{\Omega}}{\d\alpha_{m+1;lk}}\frac{\partial\alpha_{m+1;lk}}{\partial\alpha_{m}}\nonumber \\
 & \quad+\sum_{lk}\frac{\d\tilde{\Omega}}{\d\alpha_{m+2;lk}}\frac{\partial\alpha_{m+2;lk}}{\partial\alpha_{m}}.\label{eq:domega_dalpha}
\end{align}
In other words, we decompose the sum-over-paths from $\alpha_{m}$
to $\tilde{\Omega}$ in terms of shorter paths which start further
along the computational graph. Specifically, the shorter paths start
at $\alpha_{m+1}$, $\alpha_{m+2}$, and $\tilde{\mu}$, because these
are the three variables to which $\alpha_{m}$ contributes directly
(cf. Fig.~\ref{fig:autodiff_tree}). To simplify the notation, we
define the matrix,
\begin{equation}
\beta_{m}^{\ast}\equiv\d\tilde{\Omega}/\d\alpha_{m}.
\end{equation}
Note that $\alpha_{m}$ does not contribute to $\tilde{\Omega}$ for
$m\geq M$, so 
\begin{equation}
\beta_{M}=\beta_{M+1}=0.\label{eq:beta_boundary}
\end{equation}
Substitution of known quantities into Eq.~(\ref{eq:domega_dalpha})
yields the matrix recursion relation,
\begin{equation}
\beta_{m}=c_{m}^{(g)}R+2H\beta_{m+1}-\beta_{m+2},
\end{equation}
for $m=M-1$ down to $1$. 

Finally, we obtain the desired total derivatives,
\begin{align}
\frac{\d\tilde{\Omega}}{\d H^{T}} & =\sum_{m=1}^{M-1}\sum_{lk}\frac{\d\tilde{\Omega}}{\d\alpha_{m;lk}}\frac{\partial\alpha_{m;lk}}{\partial H^{T}}\nonumber \\
 & =\alpha_{0}\beta_{1}^{\dagger}+2\sum_{m=1}^{M-2}\alpha_{m}\beta_{m+1}^{\dagger}.\label{eq:domega_dh}
\end{align}

Equations~(\ref{eq:beta_boundary})–(\ref{eq:domega_dh}) are the
basis for a recursive procedure to\emph{ }calculate $\d\tilde{\Omega}/\d H^{T}$
for all $\mathcal{O}(N)$ elements $H_{ij}$ that contribute to $\tilde{\Omega}$.
The remarkable feature of reverse-mode automatic differentiation is
that the computational cost to calculate this \emph{full }gradient
scales like the cost to calculate $\tilde{\Omega}$ itself, $\mathcal{O}(NMS)$.
Note that taking finite differences with respect to each element $H_{ji}$
individually would be $\mathcal{O}(N)$ times slower than automatic
differentiation. Moreover, unlike finite differencing, automatic differentiation
is exact up to numerical accuracy.

The traditional disadvantage of reverse-mode automatic differentiation
is that it requires storage of intermediate values in the computational
graph. This is apparent in Eq.~(\ref{eq:domega_dh}), which makes
reference to matrices $\alpha_{m}$. Storing $\alpha_{m}$ for all
$m=0,\dots M-1$ would be a prohibitive space constraint. Our solution
is to store only $\alpha_{M-2}$ and $\alpha_{M-1}$ from the forward
calculation of $\tilde{\Omega}$, and then to recalculate $\alpha_{m}$
backwards, on demand, by reordering terms in Eq.~(\ref{eq:alpha}),
\begin{equation}
\alpha_{m}=2H\alpha_{m+1}-\alpha_{m+2}.
\end{equation}

\section{\label{sec:recipes}Faster Chebyshev approximation}

Equations~(\ref{eq:mu_approx}) and~(\ref{eq:alpha}) constitute
a recursive procedure to estimate the Chebyshev moments $\mu_{m}\approx\tilde{\mu}_{m}$.
We can save a factor of two in the computational cost~\cite{Weisse06}
via the Chebyshev polynomial identity,
\begin{equation}
2T_{m}(x)T_{m'}(x)=T_{m+m'}(x)+T_{m-m'}(x).
\end{equation}
The moments $\mu_{m}=\tr T_{m}(H)$ may be expressed as
\begin{align}
\mu_{2m} & =2\tr T_{m}(H)T_{m}(H)-\mu_{0},\\
\mu_{2m+1} & =2\tr T_{m}(H)T_{m+1}(H)-\mu_{1},
\end{align}
Because $H$ is Hermitian, approximate moments $\tilde{\mu}_{m}=\tr R^{\dagger}T_{m}(H)R$
may be expressed as
\begin{align}
\tilde{\mu}_{2m} & =2\tr\alpha_{m}^{\dagger}\alpha_{m}-\tilde{\mu}_{0},\label{eq:mu_fast_1}\\
\tilde{\mu}_{2m+1} & =2\tr\alpha_{m}^{\dagger}\alpha_{m+1}-\tilde{\mu}_{1},\label{eq:mu_fast_2}
\end{align}
where $\alpha_{m}\equiv T_{m}(H)R$ are computed from Eq.~(\ref{eq:alpha}).
We compute the first two moments directly, $\tilde{\mu}_{0}=\tr R^{\dagger}R$
and $\tilde{\mu}_{1}=\tr R^{\dagger}HR$, without involving $\alpha_{0}$
or $\alpha_{1}$.

As before, the free energy approximation $\tilde{\Omega}$ is given
by Eq.~(\ref{eq:omega_approx_stoch}). We require half the number
of $\alpha_{m}$ matrices by using Eqs.~(\ref{eq:mu_fast_1}) and~(\ref{eq:mu_fast_2})
instead of Eq.~(\ref{eq:mu_approx}).

In analogy with Appendix~\ref{sec:autodiff}, we transform this recursive
procedure to calculate $\tilde{\Omega}$ into one that calculates
its gradient. After a lengthy derivation\footnote{This automatic differentiation procedure is subtle because the relevant
intermediate derivatives are formally non-analytic, and make sense
only when variations to the Hamiltonian $\d H$ are constrained to
be Hermitian. } we obtain a recursion relation for matrices $\beta_{m}^{\ast}\equiv\d\tilde{\Omega}/\d\alpha_{m}$
starting from
\begin{equation}
\beta_{M/2+1}=0\quad\mathrm{and}\quad\beta_{M/2}=2c_{M-1}^{(g)}\alpha_{M/2-1},
\end{equation}
and working down to $m=1$ via,

\begin{align}
\beta_{m} & =[m>1]2c_{2m-1}^{(g)}\alpha_{m-1}+4c_{2m}^{(g)}\alpha_{m}+2c_{2m+1}^{(g)}\alpha_{m+1}\nonumber \\
 & \quad\quad+2H\beta_{m+1}-\beta_{m+2}.
\end{align}
Above we use the Iverson bracket notation,

\begin{equation}
[P]=\begin{cases}
1 & \quad\mathrm{if\,}P\\
0 & \quad\mathrm{otherwise}
\end{cases}.
\end{equation}
Our result for the gradient of approximate free energy is
\begin{align}
\frac{\d\tilde{\Omega}}{\d H^{T}} & =c_{1}^{(g)}RR^{\dagger}-\sum_{m=1}^{M/2-1}c_{2m+1}^{(g)}RR^{\dagger}\nonumber \\
 & \quad\quad+\alpha_{0}\beta_{1}^{\dagger}+2\sum_{m=1}^{M/2-1}\alpha_{m}\beta_{m+1}^{\dagger}.\label{eq:domega_dh_2}
\end{align}
The forward procedure involving Eqs.~(\ref{eq:mu_fast_1}) and~(\ref{eq:mu_fast_2})
is correct only if $H$ is Hermitian. When differentiating this procedure,
perturbations to the Hamiltonian $H\mapsto H+\d H$ should also be
Hermitian. We must explicitly symmetrize Eq.~(\ref{eq:domega_dh_2}),
the output of automatic differentiation, to get the correct final
result.

\section{\label{sec:canonical}Ensemble of fixed electron number}

So far we have been working with the grand potential $\Omega$ at
fixed chemical potential $\mu$. For numerical calculations of finite
systems, it is often preferable to work in the canonical ensemble
where $\mu$ varies according to a fixed electron number $N_{e}$.
To calculate $\mu(N_{e})$, we use the bisection method to invert
the probing estimate $N_{e}(\mu)\approx\sum_{m}c_{m}^{(f)}\tilde{\mu}_{m}$
of Eq.~(\ref{eq:ne_approx_stoch}). The coefficients $c_{m}^{(f)}$
involve an integral over the Fermi function and must be recalculated
for each trial value of the chemical potential $\mu$. The dominant
computational cost, however, is in approximating the Chebyshev moments
$\tilde{\mu}_{m}\approx\mu_{m}$. Fortunately, the moments are independent
of chemical potential $\mu$ and thus do not need to be recalculated
during the search. The bisection method is guaranteed to converge
because $\rho(x)\approx\rho_{M}(x)$ is a strictly positive approximation
when using the damping coefficients of Eq.~(\ref{eq:jackson}).

The relevant free energy in the canonical ensemble, $F(N_{e})$, is
related to the grand potential by a Legendre transform,
\begin{equation}
F(N_{e})=\Omega(\mu)+\mu N_{e}.\label{eq:canon_e}
\end{equation}
The general thermodynamic relation,
\begin{equation}
\frac{\d\Omega}{\d\mu}=-N_{e},
\end{equation}
can be verified in our context using Eqs.~(\ref{eq:free_energy}),~(\ref{eq:en_func}),
and~(\ref{eq:N_e}). The differential of $\Omega(\mu)$ at fixed
$N_{e}$ then becomes 
\begin{equation}
\d\Omega|_{N_{e}}=\frac{\d\Omega}{\d\mu}\d\mu|_{N_{e}}+\d\Omega|_{\mu}=-N_{e}\d\mu|_{N_{e}}+\d\Omega|_{\mu}.\label{eq:diff_e}
\end{equation}
We take the differential of both sides of Eq.~(\ref{eq:canon_e}),
at fixed $N_{e}$, and substitute Eq.~(\ref{eq:diff_e}) to obtain
\begin{equation}
\d F|_{N_{e}}=\d\Omega|_{\mu},\label{eq:differential_identity}
\end{equation}
Thus, in both ensembles, the derivative of the relevant free energy
with respect to the Hamiltonian yields the density matrix $f(H)$,
\begin{equation}
\left.\frac{\d F}{\d H^{T}}\right|_{N_{e}}=\left.\frac{\d\Omega}{\d H^{T}}\right|_{\mu}=f(H).
\end{equation}
In a numerical calculation, for which we approximate $\Omega\approx\tr g(H)RR^{\dagger},$
we should also substitute $N_{e}\approx\tr f(H)RR^{\dagger}$ in the
definition of $F$ in Eq.~(\ref{eq:canon_e}). This way the thermodynamic
identity of Eq.~(\ref{eq:differential_identity}) continues to hold
exactly, with or without the probing approximation.

\section{\label{sec:gradient_decay}Error analysis for gradient-based probing}

\subsection{Main result}

In Sec.~\ref{sec:density_matrix_estimation} we argued that gradient-based
probing
\begin{equation}
f(H)=\frac{\d}{\d H^{T}}\tr g(H)RR^{\dagger}+\Delta f^{\textrm{grad}}(H)
\end{equation}
yields better density matrix estimates than does direct probing. Here
we derive the asymptotic scaling of the gradient-based probing error,
$\Delta f(H)^{\text{grad}}$, valid for zero-temperature metals. Using
the fact that $(RR^{\dagger})_{ii}=1$, we can rewrite Eq.~(\ref{eq:err_autodiff})
in an explicit form,
\begin{equation}
\Delta f(H)_{ij}^{\text{grad}}=\sum_{m\neq n}\frac{\partial g(H)_{mn}}{\partial H_{ji}}\left(RR^{\dagger}\right)_{nm}.
\end{equation}
If the matrix elements of $R$ are uncorrelated and defined according
to Eq.~(\ref{eq:R_uncorrelated}), the variance of the error is
\begin{equation}
\text{var}[\Delta f(H)_{ij}^{\text{grad}}]=\frac{1}{S}\sum_{m<n}\left|\frac{\partial g(H)_{mn}}{\partial H_{ji}}\right|^{2},
\end{equation}
in analogy to Eq.~(\ref{eq:var_uncorrelated}). Note that the sum
cannot contribute any $S$ dependence, so $\Delta f(H)_{ij}^{\text{grad}}\sim S^{-1/2}$
in the uncorrelated case.

If $R$ is optimized according to Eq.~(\ref{eq:R_correlated}), we
instead have
\begin{equation}
\text{var}[\Delta f(H)_{ij}^{\text{grad}}]=\sum_{m<n}\delta_{c(m),c(n)}\left|\frac{\partial g(H)_{mn}}{\partial H_{ji}}\right|^{2},\label{eq:var_grad}
\end{equation}
in analogy to Eq.~(\ref{eq:var_correlated}). The constraint $c(m)=c(n)$
is only satisfied for orbitals $m$ and $n$ whose real-space distance
satisfies $r_{mn}\gtrsim S^{1/d}$, and thus introduces a nontrivial
dependence on $S$.

For insulators and for systems at finite temperature, $g(H)_{mn}$
decays exponentially in $r_{mn}=|\mathbf{r}_{m}-\mathbf{r}_{n}|$.
As a consequence, $\Delta f(H)_{ij}^{\text{grad}}$ decays exponentially
in $S^{1/d}$ for spatial dimension $d$. Metals at zero temperature,
however, give rise to universal power law decay. We focus our analysis
on these systems because they represent a worst-case scenario. We
saw in Eq.~(\ref{eq:g_decay}) that $g(H)_{mn}$ decays like $r_{mn}^{-(d+3)/2}$.
As we will show in Appendix~\ref{subsec:matrix_derivative}, the
corresponding matrix derivative also decays polynomially,

\begin{equation}
\frac{\partial g(H)_{mn}}{\partial H_{ji}}\sim\frac{1}{(r_{mj}r_{ni})^{\frac{d-1}{2}}(r_{mj}+r_{ni})^{2}}.\label{eqC:dgdh_asymptotics}
\end{equation}
This decay is valid when $r_{mj}=|\mathbf{r}_{m}-\mathbf{r}_{j}|$
and $r_{ni}=|\mathbf{r}_{n}-\mathbf{r}_{i}|$ are large compared to
the inverse Fermi momentum, $k_{F}^{-1}$.

When $S$ is large, we can work in the continuum limit to calculate
\begin{align}
\text{var}[\Delta f( & H)_{ij}^{\text{grad}}]\sim\nonumber \\
 & \frac{1}{S}\int\d^{d}\mathbf{r}_{m}\d^{d}\mathbf{r}_{n}\,\Theta(r_{mn}-S^{\frac{1}{d}})\left|\frac{\partial g(H)_{mn}}{\partial H_{ji}}\right|^{2}.\label{eq:var_grad_cont}
\end{align}
The Heaviside step function $\Theta(r_{mn}-S^{1/d})$ encodes the
fact that orbitals $m$ and $n$ can only contribute if separated
by distance $r_{mn}\gtrsim S^{1/d}$. The outer factor of $S^{-1}$
encodes the fact that arbitrary indices $m$ and $n$ only have the
same color with probability $S^{-1}$. This uniform probabilistic
treatment of the integrand is justified because $\partial g(H)_{mn}/\partial H_{ji}$
decays sufficiently slowly.

Our interest is estimation of \emph{local} elements $f(H)_{ij}$,
for which $r_{ij}=|\mathbf{r}_{i}-\mathbf{r}_{j}|\ll S^{1/d}$ in
the large $S$ limit. Without loss of generality, we may take $i=j=0$
in our scaling calculation. Substitution of Eq.~(\ref{eqC:dgdh_asymptotics})
yields
\begin{align}
\text{var}[\Delta f( & H)_{ij}^{\text{grad}}]\sim\nonumber \\
 & \frac{1}{S}\iint\d^{d}\mathbf{r}_{m}\d^{d}\mathbf{r}_{n}\,\frac{\Theta(r_{mn}-S^{1/d})}{(r_{m}r_{n})^{d-1}(r_{m}+r_{n})^{4}}.\label{eq:var_grad_cont_2}
\end{align}
Because $S^{1/d}$ is the only length scale in the integral, we can
perform dimensional analysis to find our final result,
\begin{equation}
\text{var}[\Delta f(H)_{ji}^{\text{grad}}]\sim S^{-(d+2)/d}.\label{eq:var_gradient_final}
\end{equation}
The behavior of $\partial g(H)_{mn}/\partial H_{00}$ as $r_{m}\rightarrow0$
and $r_{n}\rightarrow0$ is not pertinent to this scaling result.
Indeed, if we modify Eq.~(\ref{eq:var_grad_cont_2}) to constrain
$r_{m}>\sigma$ and $r_{n}>\sigma$ for some length scale $\sigma$
that satisfies $k_{F}^{-1}\ll\sigma\ll S^{1/d}$, then Eq.~(\ref{eq:var_gradient_final})
still holds.

\subsection{\label{subsec:matrix_derivative}Asymptotic decay of the energy matrix
derivative}

Here we derive the asymptotic decay of $\partial g(H)_{mn}/\partial H_{ij}$
for a model metallic Hamiltonian.

We begin with the representation of the Dirac-$\delta$ function,
\begin{equation}
\delta(x-\epsilon_{\mathbf{k}})=-\frac{1}{\pi}\mathrm{Im}\,\frac{1}{x-\epsilon_{\mathbf{k}}+i\eta}\qquad\eta\rightarrow0^{+}.\label{eq:greens_delta}
\end{equation}
This identity generalizes to a matrix equation for Hermitian $H$,
\begin{equation}
\delta(x-H)=\frac{i}{2\pi}\left[G^{+}(x)-G^{-}(x)\right].\label{eq:delta_H}
\end{equation}
The retarded/advanced Green's functions are defined as
\begin{equation}
G^{\pm}(x)=\frac{1}{x-H\pm i\eta}.
\end{equation}
We employ finite $\eta$ to regularize intermediate calculations,
with the understanding that eventually $\eta\rightarrow0^{+}$.

Equation~(\ref{eq:delta_H}) yields a differentiable representation
of the energy matrix,
\begin{equation}
g(H)=\frac{i}{2\pi}\int_{-\infty}^{+\infty}\d x\,g(x)\left[G^{+}(x)-G^{-}(x)\right].\label{eq:g_greens}
\end{equation}
We seek the derivative with respect to an arbitrary matrix element
$H_{ij}$. For any invertible operator $B$ we have $B^{-1}\partial_{\alpha}\left(BB^{-1}\right)=0$;
applying the product rule, we conclude that $\partial_{\alpha}B^{-1}=-B^{-1}(\partial_{\alpha}B)B^{-1}$.
Substituting $B^{-1}\mapsto G^{\pm}$ and $\alpha\mapsto H_{ij}$,
we find

\begin{equation}
\frac{\partial G^{\pm}(x)}{\partial H_{ij}}=G^{\pm}(x)\Delta^{ij}G^{\pm}(x),\label{eq:dyson_eq}
\end{equation}
where $\Delta^{ij}=\partial H/\partial H_{ij}$ is the matrix with
real-space elements $\Delta_{mn}^{ij}=\delta_{im}\delta_{jn}$. Equivalence
to the Dyson equation~\cite{Economou_book}, at first order in the
perturbation $\epsilon\Delta$, is apparent after expanding $\partial G^{\pm}/\partial H_{ij}\approx(G_{H+\epsilon\Delta}^{\pm}-G_{H}^{\pm})/\epsilon$.

The above identities are valid for any Hamiltonian. Now we focus on
a translation invariant Hamiltonian $H=H_{0}$ with quadratic dispersion
$\epsilon_{\mathbf{k}}=k^{2}/2$ that is filled to chemical potential
$\mu=k_{F}^{2}/2$. In momentum-space, the non-interacting Green's
functions are
\begin{equation}
G_{0}^{\pm}(x)=\int\d^{d}\mathbf{k}\,\frac{1}{x-\epsilon_{\mathbf{k}}\pm i\eta}|\mathbf{k}\rangle\langle\mathbf{k}|.
\end{equation}
The\textbf{ $\mathbf{k}$} integrals run over the first Brillouin
zone. The eigenstates $|\mathbf{k}\rangle$ have real-space representation
\begin{equation}
\langle\mathbf{r}_{i}|\mathbf{k}\rangle=\frac{1}{(2\pi)^{d/2}}e^{i\mathbf{k}\cdot\mathbf{r}_{i}}.
\end{equation}
We have assumed that the volume of the primitive cell is one.

We evaluate Eq.~(\ref{eq:dyson_eq}) at $H=H_{0}$. Inserting $\Delta^{ij}=|\mathbf{r}_{i}\rangle\langle\mathbf{r}_{j}|$,
we find matrix elements
\begin{align}
\frac{\partial G^{\pm}(x)_{mn}}{\partial H_{ij}} & =\langle\mathbf{r}_{m}|G_{0}^{\pm}(x)|\mathbf{r}_{i}\rangle\langle\mathbf{r}_{j}|G_{0}^{\pm}(x)|\mathbf{r}_{n}\rangle\nonumber \\
 & \mkern-64mu=\frac{1}{(2\pi)^{2d}}\iint\d^{d}\mathbf{k}\,\d^{d}\mathbf{k}'\,\frac{e^{i\mathbf{k}\cdot\mathbf{r}_{mi}-i\mathbf{k}'\cdot\mathbf{r}_{nj}}}{(x-\epsilon_{\mathbf{k}}\pm i\eta)(x-\epsilon_{\mathbf{k}'}\pm i\eta)}.
\end{align}
where $\mathbf{r}_{mi}=\mathbf{r}_{m}-\mathbf{r}_{i}$ and $\mathbf{r}_{nj}=\mathbf{r}_{n}-\mathbf{r}_{j}$.

Taking the derivative of both sides of Eq.~(\ref{eq:g_greens}),
we find
\begin{align}
\frac{\partial g(H)_{mn}}{\partial H_{ij}}=\frac{1}{(2\pi)^{2d}} & \int\d x\,g(x)\iint\d^{d}\mathbf{k}\,\d^{d}\mathbf{k}'\,\nonumber \\
 & \times e^{i\mathbf{k}\cdot\mathbf{r}_{mi}-i\mathbf{k}'\cdot\mathbf{r}_{nj}}W_{\mathbf{k}\mathbf{k}'}(x),
\end{align}
where, in the eventual limit that $\eta\rightarrow0^{+}$,
\begin{align}
W_{\mathbf{k}\mathbf{k}'}(x) & =-\frac{1}{\pi}\textrm{Im}\,\frac{1}{(x-\epsilon_{\mathbf{k}}+i\eta)(x-\epsilon_{\mathbf{k}'}+i\eta)}\nonumber \\
 & =\textrm{Re}\left[\frac{\delta(x-\epsilon_{\mathbf{k}})}{x-\epsilon_{\mathbf{k}'}+i\eta}+\frac{\delta(x-\epsilon_{\mathbf{k}'})}{x-\epsilon_{\mathbf{k}}+i\eta}\right].
\end{align}
The last equality follows from the identity 
\begin{equation}
\textrm{Im}(ab)=\textrm{Im}(a)\,\textrm{Re}(b)+\textrm{Im}(b)\,\textrm{Re}(a),
\end{equation}
and application of Eq.~(\ref{eq:greens_delta}), which is valid up
to irrelevant corrections for small $\eta$.

\begin{widetext}Substitution yields
\begin{align}
\frac{\partial g(H)_{mn}}{\partial H_{ij}} & =\frac{1}{(2\pi)^{2d}}\iint\d^{d}\mathbf{k}\,\d^{d}\mathbf{k}'\,e^{i\mathbf{k}\cdot\mathbf{r}_{mi}-i\mathbf{k}'\cdot\mathbf{r}_{nj}}\textrm{Re}\left[\frac{g(\epsilon_{\mathbf{k}})}{\epsilon_{\mathbf{k}}-\epsilon_{\mathbf{k}'}+i\eta}+\frac{g(\epsilon_{\mathbf{k}'})}{\epsilon_{\mathbf{k}'}-\epsilon_{\mathbf{k}}+i\eta}\right]\nonumber \\
 & =\frac{2}{(2\pi)^{d}}\int\d^{d}\mathbf{k}\,g(\epsilon_{\mathbf{k}})\left[e^{i\mathbf{k}\cdot\mathbf{r}_{mi}}\textrm{Re}\,\mathcal{G}_{0}^{+}(k,r_{nj})+e^{-i\mathbf{k}\cdot\mathbf{r}_{nj}}\textrm{Re}\,\mathcal{G}_{0}^{+}(k,r_{mi})\right].\label{eqC:derivative}
\end{align}
The non-interacting Green's function integrals are~\cite{Economou_book},

\begin{equation}
\mathcal{G}_{0}^{+}(k,r)=\frac{1}{(2\pi)^{d}}\int_{\mathbb{R}^{d}}\d^{d}\mathbf{k}'\,\frac{e^{i\mathbf{k}'\cdot\mathbf{r}}}{k^{2}-k^{\prime2}+i\eta}=\begin{cases}
-\frac{i}{2\sqrt{k^{2}+i\eta}}e^{i\sqrt{k^{2}+i\eta}\,r} & d=1\\
-\frac{i}{4}H_{0}^{(1)}(\sqrt{k^{2}+i\eta}\,r) & d=2\\
-\frac{1}{4\pi r}e^{i\sqrt{k^{2}+i\eta}\,r} & d=3
\end{cases},\label{eqC:Greens}
\end{equation}
where $H_{0}^{(1)}$is the Hankel function of the first kind. Above
we have extended the integration domain from the first Brillouin zone
to $\mathbb{R}^{d}$; this continuum limit (lattice parameter $a\rightarrow0$)
is valid when $ak_{F}\ll1$.

We seek to evaluate Eq.~(\ref{eqC:derivative}) in the limit $\eta\rightarrow0^{+}$.
Note that $\mathcal{G}_{0}^{+}(k=0,r)$ is singular in this limit
for $d=1$ and $d=2$. In one dimension, we use the fact that $\textrm{Im}\,(k^{2}+i\eta)^{-1/2}=-\pi\delta(k)/2$.
In two dimensions, the $(k^{2}+i\eta)^{-1/4}$ singularity at $\mathcal{G}_{0}^{+}(k=0,r)$
will be scaled by a factor of $kJ_{0}(kr)\sim\sqrt{k}$ in the integrand,
and can be ignored. Thus, when $\eta\rightarrow0^{+}$, we employ
\begin{equation}
\textrm{Re}\,\mathcal{G}_{0}^{+}(k,r)=\begin{cases}
-\frac{\pi}{4}\delta(k)+\frac{\sin(kr)}{2k} & d=1\\
\frac{1}{4}Y_{0}(kr) & d=2\\
-\frac{\cos(kr)}{4\pi r} & d=3
\end{cases},
\end{equation}
where $Y_{0}(kr)$ is the Bessel function of the second kind. Substitution
into Eq.~(\ref{eqC:derivative}) with $g(\epsilon_{\mathbf{k}})=(k^{2}-k_{F}^{2})\Theta(k_{F}-k)/2$
yields
\begin{align}
\left.\frac{\partial g(H)_{mn}}{\partial H_{ij}}\right|_{d=1} & =\frac{k_{F}^{2}}{4}+\frac{1}{2\pi}\int_{0}^{k_{F}}\d k\,(k^{2}-k_{F}^{2})\frac{\sin\left[k(r_{mi}+r_{nj})\right]}{k}\approx\frac{\sin\left[k_{F}(r_{mi}+r_{nj})\right]}{\pi(r_{mi}+r_{nj})^{2}},\label{eq:dg_dh_1}\\
\left.\frac{\partial g(H)_{mn}}{\partial H_{ij}}\right|_{d=2} & =\frac{1}{4}\int_{0}^{k_{F}}\d k\,k(k^{2}-k_{F}^{2})\left[J_{0}(kr_{mi})Y_{0}(kr_{nj})+J_{0}(kr_{nj})Y_{0}(kr_{mi})\right]\approx k_{F}\frac{\cos\left[k_{F}(r_{mi}+r_{nj})\right]}{\sqrt{r_{mi}r_{nj}}(r_{mi}+r_{nj})^{2}},\\
\left.\frac{\partial g(H)_{mn}}{\partial H_{ij}}\right|_{d=3} & =-\frac{1}{8\pi^{3}r_{mi}r_{nj}}\int_{0}^{k_{F}}\d k\,k(k^{2}-k_{F}^{2})\sin\left[k(r_{mi}+r_{nj})\right]\approx-k_{F}^{2}\frac{\sin\left[k_{F}(r_{mi}+r_{nj})\right]}{4\pi^{3}r_{mi}r_{nj}(r_{mi}+r_{nj})^{2}}.\label{eq:dg_dh_3}
\end{align}

\end{widetext}The final three approximations are asymptotically valid
when $r_{mi}$ and $r_{nj}$ are large compared to $k_{F}^{-1}$.
We conclude, in all dimensions $d$, that the matrix derivative decays
as
\begin{equation}
\frac{\partial g(H)_{mn}}{\partial H_{ij}}\sim\frac{1}{(r_{mi}r_{nj})^{\frac{d-1}{2}}(r_{mi}+r_{nj})^{2}}\label{eqC:dgdh}
\end{equation}
for large $r_{mi}$ and $r_{nj}$. This power law decay is universal
and we have verified it numerically in the context of simple tight-binding
models.

\bibliographystyle{apsrev4-1}
\bibliography{refs}

\end{document}